\documentclass[twocolumn,prb,aps,citeautoscript,showpacs,footinbib,amsmath,amssymb,superscriptaddress,longbibliography]{revtex4-1}
\usepackage{array}
\usepackage{graphicx}
\usepackage{dcolumn}
\usepackage{color}
\usepackage{bm}
\usepackage{natbib}
\usepackage{hyperref}
\usepackage{amsmath,amssymb,amsfonts,bbold}
\usepackage[normalem]{ulem}
\usepackage{float}

\usepackage{tikz}

\newcommand{\beq}{\begin{equation}}
\newcommand{\eeq}{\end{equation}}
\newcommand{\beqa}{\begin{eqnarray}}
\newcommand{\eeqa}{\end{eqnarray}}

\begin{document}

\title{Magnetic domain structure of epitaxial Gd films grown on W(110)}

\author{Patrick H\"{a}rtl}
	\email[corresponding author: ]{patrick.haertl@physik.uni-wuerzburg.de}
	\address{Physikalisches Institut, Experimentelle Physik II, 
	Universit\"{a}t W\"{u}rzburg, Am Hubland, 97074 W\"{u}rzburg, Germany}
\author{Markus Leisegang}
	\affiliation{Physikalisches Institut, Experimentelle Physik II, 
	Universit\"{a}t W\"{u}rzburg, Am Hubland, 97074 W\"{u}rzburg, Germany} 
\author{Matthias Bode} 
	\address{Physikalisches Institut, Experimentelle Physik II, 
	Universit\"{a}t W\"{u}rzburg, Am Hubland, 97074 W\"{u}rzburg, Germany}	
	\address{Wilhelm Conrad R{\"o}ntgen-Center for Complex Material Systems (RCCM), 
	Universit\"{a}t W\"{u}rzburg, Am Hubland, 97074 W\"{u}rzburg, Germany}    
	   


\date{\today}

\begin{abstract}
We present a detailed real-space spin-polarized scanning tunneling microscopy (SP-STM) study 
of the magnetic domain structure of Gd(0001) films epitaxially grown on W(110). 
To find optimal preparation conditions, the influence of the substrate temperature during deposition 
and of the post-growth annealing temperature was investigated.  
Our results show that the lowest density of surface defects, such as step edges as well as screw and edge dislocations, 
is obtained for room temperature deposition and subsequent annealing at 900\,K.   
SP-STM data reveal small-sized magnetic domains at lower annealing temperatures, 
evidently caused by pinning at grain boundaries and other crystalline defects.  
The coverage-dependent magnetic domain structure of optimally prepared Gd films was systematically investigated. 
For low coverage up to about 80 atomic layers (AL) we observe $\mu$m-sized domains separated by domain walls 
which are oriented approximately along the $[1{\bar 1}0]$ direction of the underlying W substrate.  
Above a critical film thicknesses $\Theta_{\text{crit}} \approx (100 \pm 20)$\,AL, we identify stripe domains, 
indicative for a spin reorientation transition from in-plane to out-of-plane.  
In agreement with existing models, the periodicity of the stripe domains increases 
the further the coverage exceeds $\Theta_{\text{crit}}$.
While the orientation of the stripe domains is homogeneous over large distances just above $\Theta_{\text{crit}}$, 
we find a characteristic zig-zag pattern at $\Theta \gtrsim 200$\,AL and irregular stripe domains beyond 500\,AL.   
Intermediate minima and maxima of the magnetic signal indicate the nucleation of branching domains.  
The results are discussed in terms of various contributions to the total magnetic energy, 
such as the magneto-crystalline, the magneto-static, and the magneto-elastic energy density.
\end{abstract}


\maketitle


\section{Introduction}
\label{sect:Introduction}
\vspace{-0.3cm}
The functionality of numerous technical devices relies on magnetic thin films.
For example, they are used as positioning or speed sensors \cite{Ripka2010} and magnetic hard disk drives 
still represent the backbone of present mass data storage applications \cite{Bhat2018}.
The progress we have witnessed in the past decades required a thorough understanding of the physical processes 
which determine the properties of thin magnetic films, such as the saturation magnetization, 
the remanent magnetization, or the coercive field.  
In polycrystalline or granular materials, however, the intricate interplay between structural defects 
and magnetic domain walls impedes a clear identification of the underlying physical processes.  

In this context, thin films grown epitaxially on single crystalline and highly pure substrates offer an invaluable playground 
to study the impact structural properties have on the magnetization behavior.  
Surface-adapted \cite{JMMM53.L295} and surface-sensitive \cite{SWT2000} magnetic measurement techniques 
allow for a detailed understanding, especially if combined with microscopic imaging 
methods \cite{Bode2003,SW2008,RS2010,McCord_2015,HGW2017,Kohashi2018}.  
In particular, thin $3d$ transition metal (TM) films deposited on the surfaces 
of noble or refractory metals were intensively studied.  
For example, these investigations provided consistent pictures of the onset of long-range order 
at island coalescence \cite{PhysRevLett.73.898,PhysRevB.60.7379}, 
of film thickness-dependent spin reorientation transitions  \cite{PhysRevLett.69.3385,PhysRevLett.69.3831,PhysRevLett.93.117205},
and of the development of a uniaxial anisotropy on vicinal, highly stepped surfaces \cite{PhysRevLett.68.839,PhysRevLett.77.2570}. 
Furthermore, the capability of imaging magnetic surfaces with atomic spin resolution 
by spin-polarized scanning tunneling microscopy (SP-STM) \cite{Bode2003} led to the discovery 
of highly complex spin structures, such as ferro- and antiferromagnetic spin cycloids \cite{Bode2007,Ferriani2008} 
and two-dimensional magnetic skyrmions \cite{Heinze2011}.  

In contrast, very little is known about the magnetic domain structures of thin rare-earth metal (REM) films.   
Recent real-space imaging studies are limited to Dy \cite{Krause2006,PhysRevB.76.064411,PhysRevB.80.241408}, 
Tb\cite{Prieto2016}, and Nd \cite{Kamber2020} films deposited on W(110).  
Given the important role REMs play in permanent magnets and considering 
their fundamentally different magnetic coupling mechanism,
which relies on the Ruderman-Kittel-Kasuya-Yosida (RKKY) interaction 
rather than the direct exchange at work in $3d$ TMs, this is very surprising.  
Potentially, this lack of high-spatial resolution magnetic domain studies is related to the extreme reactivity of REMs
which not only imposes high standards on the cleanliness of the substrate and the evaporant, 
but also on the vacuum conditions during deposition \cite{Getzlaff1999,PhysRevB.76.113410}.  

\begin{figure}[b]
	\centering
	\includegraphics[width=\linewidth]{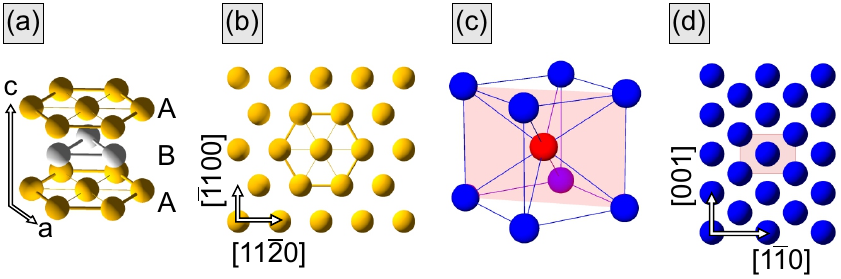}
			\caption{(a) The hexagonal close-packed A/B/A stacking model for Gadolinium including the a- and c-axis. 
			(b) The corresponding $(0001)$ plane of this structure 
			with the crystallographic ${[11\bar{2}0]}$ and ${[\bar{1}100]}$ directions. 
			(c) Body-centered cubic unit cell of Tungsten with (d) the atomic structure of the $(110)$ plane 
			and the ${[1\bar{1}0]}$ and ${[001]}$ directions.}
		\label{Fig:crystalmodels}
\end{figure}
Gadolinium (Gd) is considered to be the prototype ferromagnet among the lanthanides with localized magnetic moments. 
Bulk Gd is ferromagnetic with a Curie temperature ${T_{\text{C}}\,=\,293\,\text{K}}$ \cite{Baberschke1987}. 
Since its $4f$ shell is half filled, Hund coupling results in a magnetic $4f$ spin moment of $7\,{\mu_{\rm B}}$ 
and a vanishing orbital moment with an almost spherical charge distribution.  
Accordingly, in comparison with the other REMs, the magneto-crystalline anisotropy energy density of bulk Gd is very low.   

Gd crystallizes in the hexagonal close-packed (hcp) crystal structure  
with an alternating sequence of A/B/A layers, see Fig.\,\ref{Fig:crystalmodels}(a,b). 
With a ${c/a}$ ratio of 1.59, the hcp lattice structure of bulk Gd deviates somewhat from the ideal value, ${c/a\,=\,1.633}$.  
This causes an anisotropic dipole-dipole interaction which, if considered in isolation, 
would result in a magnetization pointing along the c-axis. 
In combination with the above-mentioned small magneto-crystalline anisotropy, it results in an 
easy axis which---at low temperatures relevant for this study---is rotated by about $30^{\circ}$ away from the c-axis \cite{Corner1975}.

The magnetic properties of Gd thin films epitaxially grown on the (110) surface 
of refractory body-centered cubic (bcc) W single-crystals have been investigated in numerous studies. 
The epitaxial growth relation beweeen these two materials is Gd(0001) $\parallel$ W(110) 
and ${\text{Gd}[11\bar{2}0] \parallel \text{W}[1\bar{1}0]}$ \cite{Weller1986}.
It was recognized at an early stage that both the film thickness \cite{Farle1993} 
and the annealing temperature \cite{PhysRevB.45.503,Tober1996} significantly affect the magnetic properties. 
The observed enhancement of the ac susceptibility $\chi_{\rm ac}$ was interpreted in terms of misfit dislocations 
which are annealed at elevated temperature, whereby the temperature required to optimize the height and width 
of the $\chi_{\rm ac}$ peak increased with increasing film thickness \cite{PhysRevB.45.503}.

Susceptibility measurements also resulted in the identification 
of a spin reorientation transition (SRT) from in-plane at low film thickness $\Theta$ to out-of-plane magnetized stripe domains 
for $\Theta \geq {\Theta_{\text{crit}} = 40\,\text{nm}}$ \cite{PhysRevB.50.6457,BergerJMMM,PhysRevB.52.1078}. 
This transition was explained with the competition of two energy contributions, i.e., 
the stray field or magneto-static energy and the uniaxial magneto-crystalline anisotropy energy. 
Whereas the former dominates for thin films, 
the latter becomes more relevant for thick films and leads to a rotation of the magnetization towards the c-axis, 
i.e., perpendicular to the surface plane of Gd(0001)/W(110) \cite{BergerJMMM,PhysRevB.52.1078}. 

A lively discussion developed around the question whether or not the topmost surface layer 
of Gd(0001) films on W(110) possesses extraordinary magnetic properties, 
such as an enhanced surface Curie temperature \cite{Weller1985,PhysRevLett.71.444}
or an imperfect or even antiferromagnetic coupling to the bulk \cite{Mulhollan1992,Li_1993}. 
In this context a spin-split $d_{z^2}$--like surface state of Gd(0001) was intensively 
investigated \cite{PhysRevB.49.7734,Weschke1996,FEDOROV1998,GETZLAFF1998,PhysRevLett.83.3017}, 
essentially rebutting any surface-related magnetic anomaly.  
The well-defined spin polarization of the occupied majority and the unoccupied minority part 
of this surface state was later used to establish the spectroscopic mode 
of spin-polarized scanning tunneling microscopy (SP-STM) \cite{Bode1998,APL1999}.

Here we report on a systematic SP-STM investigation of the film thickness-dependent magnetic domain structure 
of Gd(0001) films epitaxially grown on W(110). 
To identify optimal growth conditions both the substrate temperature during Gd deposition 
and the post-deposition annealing temperature were modified. 
We find that room temperature deposition results in the smoothest films 
with the lowest density of surface defects, such as step edges, screw and edge dislocations.
Spin-resolved measurements clearly show that annealing Gd films at a temperature of 600\,K 
is insufficient to release stacking faults from the as-grown films, 
resulting in relatively small domains, probably due to domain wall pinning. 
Only upon annealing to 900\,K large domains with straight domain walls were observed.  
A coverage-dependent SP-STM study of Gd/W(110) in the coverage range between 20\,AL and 600\,AL 
finds a spin-reorientation transitions (SRT) from in-plane to out-of-plane 
at a critical coverage $\Theta_{\text{crit}} \approx (100 \pm 20)$\,AL, in excellent agreement 
with earlier findings based on susceptibility measurements \cite{BergerJMMM,PhysRevB.52.1078}.  
Stripe domains are imaged for $\Theta > \Theta_{\text{crit}}$. 
The high spatial resolution of SP-STM allows for an investigation of the stripe periodicity.  
While we find that the magnetic stripe domains are homogeneously tilted by either $+30^{\circ}$ or $-30^{\circ}$ 
with respect to the W$[001]$ direction just above $\Theta_{\text{crit}}$, 
a zig-zag pattern with an alternating stripe orientation is found between 200\,AL and 375\,AL.  
In the same coverage regime we find evidence for the formation of branching domains.
At even higher Gd coverage irregular stripe domains are found.
The observed magnetic domain structures are discussed in terms of competing effects, 
i.e., the magneto-crystalline, magneto-static, and magneto-elastic contributions to the total energy.

\section{Experimental setup and procedures}
\label{sect:Experiment}
\vspace{-0.3cm}

The experiments were performed in a two-chamber ultra-high vacuum (UHV) system 
with a base pressure ${p\,\le\,5\times 10^{-11}\,\text{mbar}}$. 
A preparation chamber facilitates tip and sample preparation by electron beam heating to temperatures well above 2300\,K. 
Variable leak valves allow for the dosing of high-purity gases.  
The $\text{W}(110)$ single crystal was initially cleaned in an oxygen atmosphere by a two-step-flashing procedure 
with consecutive low-temperature ($T_{\rm sample} \approx 1200$\,K) 
and high-temperature ($T_{\rm sample} \gtrsim 2200$\,K) flashes \cite{Zakeri2010}. 
Flash temperatures were measured with an optical pyrometer (Ircon Ultimax UX-20P)
at an emissivity $\epsilon = 0.33$ and an estimated accuracy of $\pm 100$\,K.
The cleanliness of the substrate was confirmed by a sharp low-energy electron diffraction (LEED) pattern 
with a low background intensity, as will be shown further below, and by scanning tunneling microscopy (STM).

Distilled Gd lumps (MaTeck, purity $99.99\,\%$) were melted in a Mo crucible 
and deposited by electron-beam evaporation.   
The {\em absolute} deposition rate is determined to $(19.8\,\pm\,3.7)$\,atomic layers (AL) per minute \cite{SupplMat}, 
where the error margin corresponds to a statistical uncertainty due to the limited scan range of our STM images.
Therefore, we expect a much lower {\em relative} error between samples with different Gd coverage, 
as they are achieved by a simple scaling of the deposition time.  
During deposition and LEED experiments, the sample was clamped to a manipulator 
which can either be cooled by liquid nitrogen, resulting in a sample temperature ${T_{\rm sample} \gtrsim 120\,\text{K}}$, 
or heated by a filament (${T_{\rm sample} \lesssim 1000\,\text{K}}$). 
On the manipulator the sample temperature is measured by a thermocouple attached to one side of the sample slot. 
Comparison of temperature readings taken with the optical pyrometer and the thermocouple 
in the temperature range between 590\,K and 890\,K agree within $\approx 50$\,K.

Upon preparation, the samples were transferred into the STM chamber. 
This chamber includes a cryostat cooled with liquid helium which houses a home-built low-temperature STM.  
It is operated at a base temperature ${T_{\text{STM}}\approx 4.5\,\text{K}}$.  
Measurements were performed with electro-chemically etched W tips. 
All images were taken in the constant-current mode with the bias voltage ${U_{\text{bias}}}$ applied to the sample. 
Tunneling d$I$/d$U$ spectra and maps were obtained by modulating the bias voltage 
at a frequency $f_{\rm mod} = 5.309$ kHz, i.e., well above the cut-off frequency of the feedback loop, 
and detecting the resulting amplitude of the current modulation with a lock-in amplifier.  
Typical amplitudes of the bias voltage modulation were $U_{\rm mod} = 10$\,mV.  

For magnetically sensitive spin-polarized STM measurements, 
the tip was dipped several nanometers into the Gd surface 
or gently pulsed at voltages ${U_{\text{bias}}\,\geq\,\pm 4\,\text{V}}$.
As stated in more detail in Sec.\,\ref{subsec:Mag}, this preparation procedure reliably resulted in probe tips which gave a strong tunneling magneto-resistance (TMR) contrast 
in differential conductivity d$I$/d$U$ images, especially around $U_{\text{bias}} \approx -700$\,mV. 
Furthermore, it has the advantage that it comes with a much lower experimental effort 
than the preparation of thin film tips described in earlier publications \cite{Bode2003}, 
which required a high-temperature treatment and subsequent film deposition and annealing.
At the same time, however, it has the disadvantage that the tip magnetization direction is probably 
not aligned along the in-plane or out-of-plane direction but canted relative to the sample surface. 
For thin film tips, in contrast, the magnetization direction 
can be predicted with reasonable confidence\cite{Bode2003} and verified 
by scanning suitable test samples, for an example see Fig.\,S3 in Ref.\,\onlinecite{NC}. 
Our preference for a swift magnetic tip preparation technique rather than a much slower method 
with a more reliable quantization axis is a consequence of the focus of this study, 
which is on the systematic investigation of a broad variety of preparation parameters, 
such as the substrate temperature during film deposition, the post-growth annealing temperature, and the Gd film thickness.  
The sheer number of experimental runs excluded alternative tip preparation procedures. 
As a result, the easy axis of the domain or the magnetic orientation of domain walls 
cannot be deduced directly from our SP-STM results. 
Instead, our interpretation also relies on comparison with published results 
obtained with spatially averaging experimental techniques\cite{Weller1986,BergerJMMM} 
and the comparison with domain pattern observed on sample systems other than Gd.
Furthermore, the rather large amount of magnetic material in Fe-coated tips 
often led to a modification of the Gd domain patterns, as observed previously with other sample systems \cite{AFM_ProbeTips}. 
Our experience shows that magnetic tips prepared by dipping into the Gd film are much less invasive, 
probably because the effective amount of material is much smaller.

\begin{figure*}[t]
	\centering
	\includegraphics[width=\linewidth]{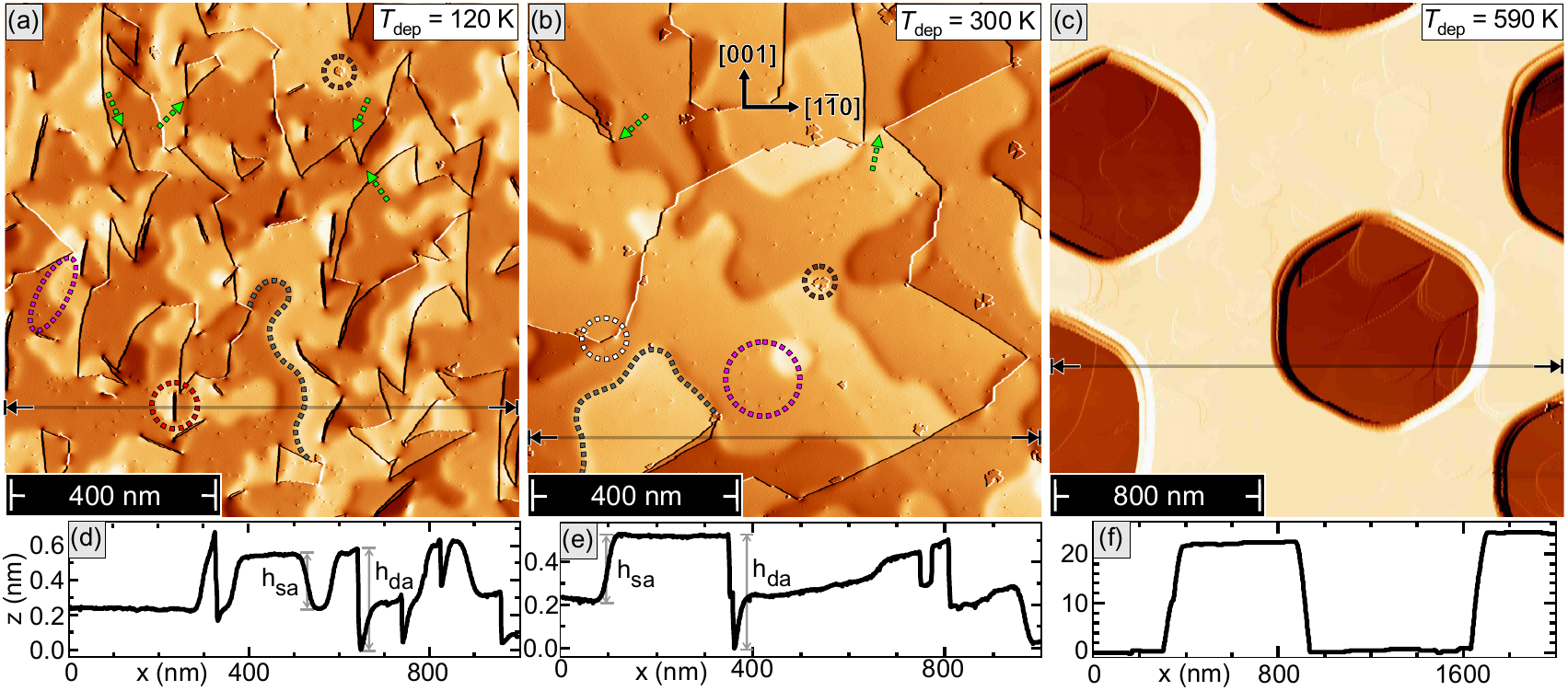}
			\caption{STM topographic images for ${80\,\text{AL}}$ thick Gd films on W(110) deposited at substrate temperatures of
			 (a) ${T_{\text{dep}}\,=\,120}$\,K, (b) $300$\,K, and (c) $590$\,K. 
			 Upon deposition, the films were annealed at ${T_{\text{ann}}\,\geq\,900\,\text{K}}$ for five minutes. 
			 The height profiles below each panel were taken along the respective lines. 
			 Scan parameters: ${U_{\text{bias}} = -700\,\text{mV}}$ and ${I_{\text{set}} = 1\,\text{nA}}$.} 
		\label{Fig:depositiontemp}
\end{figure*}

For better visibility of our STM data, the $z$-signal recorded in the topographic constant-current image 
was augmented by its derivative with respect to the fast scan direction, d$z /$d$x$. 
This image processing suggests to the observer a topography image 
that is illuminated by an invisible light source from the left.
Depending on the maximal corrugation, the exact mixing ratio of $z$ and d$z /$d$x$ varies for different images of this study.
As a result, the color-code cannot directly be interpreted as a height information. 
Wherever necessary, line profiles will be presented to allow for a quantitative assessment. 
Details of the STM image processing can be found in the supplementary material \cite{SupplMat}. 

\section{Results}
\label{sec:Experiment} 
\vspace{-0.3cm}
\subsection{Structural properties}
\label{subsec:Struc}
\vspace{-0.3cm}
\noindent
{\it Deposition temperature} --- In the literature the effect of the W(110) substrate temperature 
on the Gd film quality has been discussed to some extent. 
While the majority of studies \cite{Aspelmeier1994,Getzlaff1999,Bode1998,Getzlaff1999,GETZLAFF1998,
PhysRevLett.83.3017,Bode1998,paschen1993magnetic,PhysRevB.45.503} deposit at about 300\,K, 
some report on enhanced \cite{Tober1996,Mulhollan1992,Weller1986} 
or reduced temperatures \cite{Starke:PrivComm}.  
In order to identify the optimal growth conditions for later magnetic domain studies, 
we deposited Gd films with a thickness of $80$\,AL onto the W(110) substrate held at various temperatures.
Fig.\,\ref{Fig:depositiontemp} shows STM overviews of samples grown at 
(a) ${T_{\text{dep}}\,=\,120}$\,K, (b) ${300}$\,K, and (c) ${590}$\,K. 
Upon film growth, all samples were post-annealed on an e-beam stage at ${T_{\text{ann}}\,\geq\,900\,\text{K}}$ for five minutes. 
Height profiles measured along the black transparent lines marked in (a)-(c) are presented in Fig.\,\ref{Fig:depositiontemp}(d)-(e), respectively.

The STM topography image of the sample grown at a reduced substrate temperature, Fig.\,\ref{Fig:depositiontemp}(a),  
is dominated by numerous short step edges which often converge under an acute angle with neighboring step edges, 
thereby forming a zig-zag pattern of V-shaped single-atomic step edges with double screw dislocations 
at the joints, exemplary marked by green arrows. 
The typical length of these step edges is several ten up to a few hundred nanometers 
and the density of the double screw dislocations amounts to about $(172 \pm 13)$ per $\mu{\rm m}^2$.\footnote{
	Defect densities have been determined on the basis of representative STM images with a scan range 
	of $1\,\mu{\rm m} \times 1\,\mu{\rm m}$ recorded at a $1024 \times 1024$ pixel resolution. 
	Since the number of events $N$ measured within the randomly chosen sampling area 
	follows a Poisson distribution, the error bar given is based on the standard deviation, $\sqrt{N}$. }  
Occasionally, two single-atomic step edges emerge and terminate at the same location and run almost in parallel, 
thereby forming an isolated pair which closely resembles a double step edge.
An example is marked by a red dashed circle.  
These two features, i.e., step edges with a height in close agreement with a single-atomic ($h_{\rm sa} = 289$\,pm) 
and double-atomic step edges ($h_{\rm da} = 578$\,pm), can also be recognized 
in the line section presented in Fig.\,\ref{Fig:depositiontemp}(d) as sharp transitions. 
Furthermore, numerous point-like indentations (pink dashed circle) 
with a density of about $(341\pm 18)$ per $\mu{\rm m}^2$, 
and smoothed-out step edges (grey dashed line) can be recognized, whose origin will be explained later on. 
We would like to note that a few sharply delimited holes with an apparent depth 
of about ${(140\pm 20)}$\,pm can be recognized, one of which is marked by a brown dashed circle.  
It has been shown in previous studies that these holes are caused by the local adsorption of hydrogen,
resulting in the quenching of the $d_{z^2}$-like Gd(0001) surface state \cite{Getzlaff1999}.  
These holes cover less than 0.1\% of the total surface area, thereby corroborating the excellent cleanliness of the surface.

\begin{figure*}[t]
	\centering
	\includegraphics[width=\textwidth]{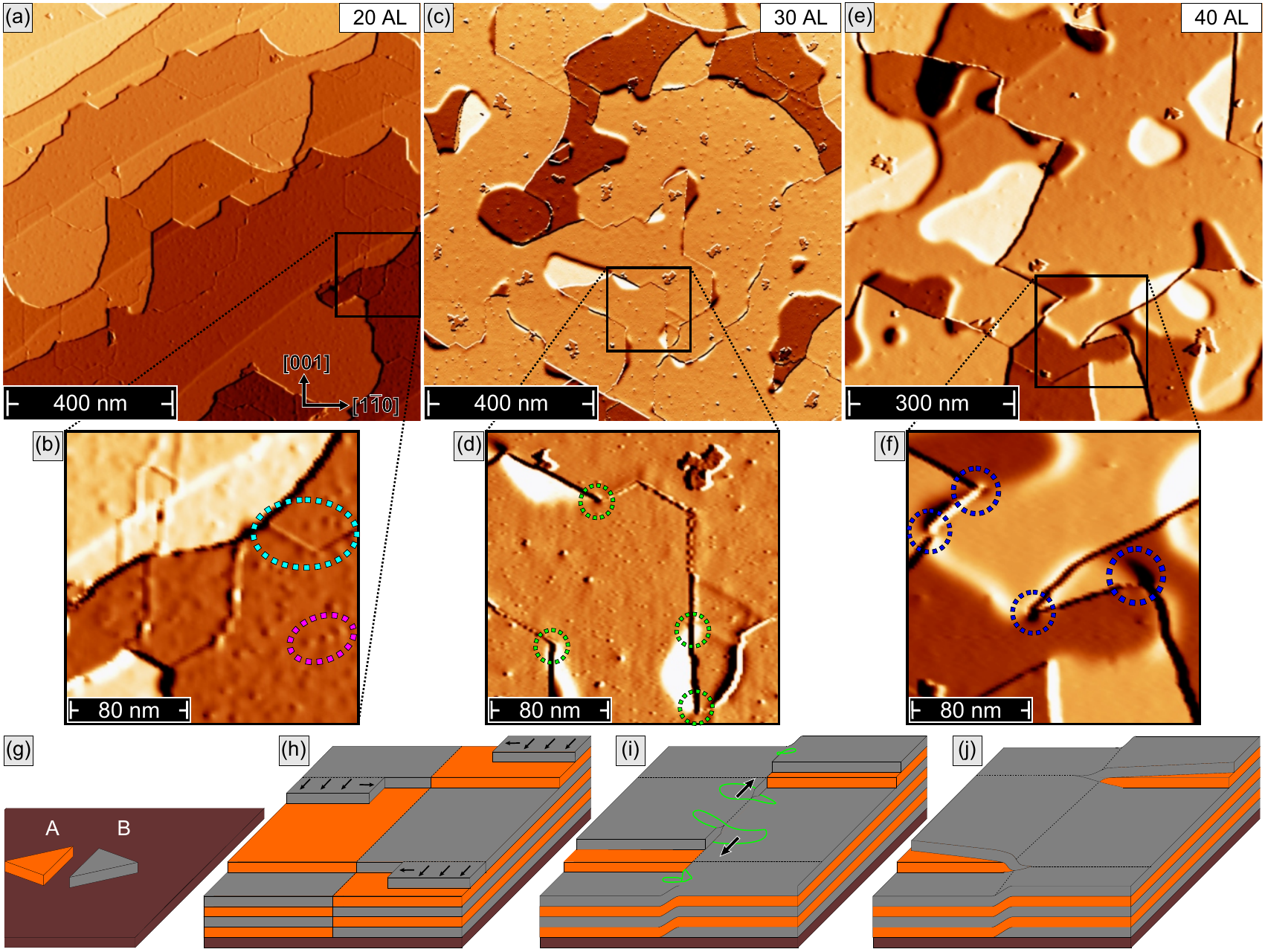}
			\caption{(a), (c), (e) STM topography images of the thickness-dependent surface structures 
			of Gd films on W(110) for (a) 20\,AL, (b) 30\,AL, and (c) 40\,AL. 
			Zoomed-in data are presented in (b), (d), (f) to highlight characteristic structural defects. 
			Scan parameters $U_{\text{bias}} = -700$\,mV, $I_{\text{set}} = 1$\,nA.
			(g-j) Model explaining the formation of structural stacking faults and their relaxation via glide dislocations. 
			Initially, (g) A- or B-stacked islands nucleate which 
			(h) lead to structural grain boundaries as the film closes (arrows indicate film growth direction).  
			(i) The resulting film stress is relaxed via the formation of screw dislocations 
			which move with increasing film thickness (indicated by arrows)
			and eventually (j) form characteristic double screw dislocations. 
			\label{Fig:layer_thickness}}
\end{figure*}

The sample prepared at room temperature, Fig.\,\ref{Fig:depositiontemp}(b), exhibits a much lower density 
of these structural features and, therefore, the overall surface morphology appears much smoother. 
For example, the density of the V-shaped single-step edges is reduced 
to about $(31 \pm 6)$ per $\mu{\rm m}^2$, i.e., a reduction of about 82\% 
as compared to the low-temperature grown film presented in Fig.\,\ref{Fig:depositiontemp}(a).  
A similar reduction is also identified for the point-like indentations (pink dashed circle) 
to about $(202 \pm 14)$ per $\mu{\rm m}^2$, which is about 41\,\% less than for the cold evaporation. 
The density of smoothed step edges (indicated by a grey dashed line) is also reduced by a similar percentage. 
The height profile presented in Fig.\,\ref{Fig:depositiontemp}(e) confirms these observations.  

The Gd film grown at ${T_{\text{dep}} = 590}$\,K shows a completely different surface morphology, see Fig.\,\ref{Fig:depositiontemp}(c). 
Here we find large hexagonal vacancy islands with depths ${h \geq 20}$\,nm 
and typical side lengths of about 400\,nm [see also the line profile in Fig.\,\ref{Fig:depositiontemp}(f)]. 
Obviously, the film thickness is no longer homogeneous but at the edge to island formation, 
in general agreement with similar observations made for thinner films \cite{Tober1996}. 

These results indicate that the optimal substrate temperature for obtaining homogeneous Gd films 
with a minimal density of structural defects is $\approx 300$\,K.   
In contrast, elevated or lowered temperatures result in surface textures with more structural defects
which potentially act as pinning centers for magnetic domain walls and are, therefore, 
unsuitable for imaging the intrinsic domain structure of Gd/W(110) films.   
A similar behavior with respect to the annealing temperature will be presented in Sec.\,\ref{subsec:Mag} below.

\noindent
{\it Growth } --- Figure~\ref{Fig:layer_thickness} summarizes the main processes 
important to understand the structural features of Gd(0001) films grown on W(110).  
All samples were deposited at room temperature and post-annealed at $T_{\rm anneal} \geq 900$\,K for 5\,min.
Fig.\,\ref{Fig:layer_thickness}(a) shows an overview of the terrace-and-step structure of a 20\,AL film.  
The overgrown step edges of the underlying W substrate can clearly be identified. 
The zoom-in image reveals that the terraces are not perfectly flat but exhibit trenches which are about ${(30\pm 5)}$\,pm deep.   
One example is marked by a cyan circle in Fig.\,\ref{Fig:layer_thickness}(b).  
Furthermore, numerous point-like indentations (pink circle) can be recognized.  

When increasing the film thickness to 30\,AL, Fig.\,\ref{Fig:layer_thickness}(c), 
the film texture changes considerably, 
as the trenches become deeper (up to 170 -- 240\,pm). 
Furthermore, they are more asymmetric and---in some segments---have even converted into full-fledged step edges. 
Some positions where the transition from an asymmetric trench into a step edge is observed, 
are marked by green dashed circles in the higher magnification image presented in Fig.\,\ref{Fig:layer_thickness}(d).  
This transition not only results in a screw dislocation, but also in the appearance of smoothed-out step edges 
which appear as white bumps in Fig.\,\ref{Fig:layer_thickness}(d).
At an even higher film thickness of 40\,AL, see Fig.\,\ref{Fig:layer_thickness}(e), 
the surface morphology is dominated by smoothed steps and double-screw dislocations, 
some of which are marked by blue dashed circles in the zoom-in in Fig.\,\ref{Fig:layer_thickness}(f). 
These features have completely replaced the trenches observed at lower coverage, which no longer can be found.  

Overall, the thickness-dependent changes observed in Fig.\,\ref{Fig:layer_thickness}(a)-(e) for Gd/W(110) 
show close resemblance to earlier results of Dy films grown on W(110) reported by Krause \emph{et al} \cite{Krause2006}.
In this report it was shown by atomic resolution data that point-like indentations 
similar to those we observed in Figs.\,\ref{Fig:depositiontemp}(a) and \ref{Fig:layer_thickness}(b)
are caused by edge dislocations, i.e., an additional semi-infinite plane which releases misfit-related strain.  
Furthermore, Krause and co-authors proposed a general model \cite{Krause2006}, 
which is reproduced in Fig.\,\ref{Fig:layer_thickness}(g)-(j), 
that describes the evolution of the rare-earth metal (REM) film morphology on bcc(110) surfaces.  
The first atomic layer of heavy REMs epitaxially grown on W(110) single crystals 
reveals a heavily distorted hexagonal lattice \cite{Kolaczkiewicz1986,Weller1986}. 
The nucleation of A- and B-stacked islands, schematically represented 
in Fig.\,\ref{Fig:layer_thickness}(g), starts with the second atomic layer \cite{GETZLAFF1998}. 
These islands grow to patches which maintain their respective stacking order, i.e., A/B/A or B/A/B, 
even after coalescence when a continuous film is formed at relatively low coverage. 
As indicated by the differently colored surface layers in Fig.\,\ref{Fig:layer_thickness}(h), however, 
the film is not translation invariant at the grain boundary between two of these patches, 
resulting in structural domain boundaries which we observe as trenches in Fig.\,\ref{Fig:layer_thickness}(a) and (b).  

Since the energy associated with the grain boundary increases with increasing film thickness, 
relaxation processes set in at a critical thickness, Fig.\,\ref{Fig:layer_thickness}(i). 
They lead to so-called partial screw dislocations which shift differently stacked patches 
by half a lattice constant along the c-axis, i.e., perpendicular to the film plane,
and result in a film with continuous A- and B-stacked layers.  
However, this process is accompanied by a significant surface buckling which in our data 
appears as smoothed-out step edges, cf.\ Figs.\,\ref{Fig:depositiontemp}(a), (b) and \ref{Fig:layer_thickness}(d).  
Obviously, for Gd films on W(110) this process starts at a coverage somewhere between 20 and 30\,AL.
As indicated by black arrows in Fig.\,\ref{Fig:layer_thickness}(i), these partial screw dislocations 
move with increasing film thickness until an adjacent pair merges, see Fig.\,\ref{Fig:layer_thickness}(j).  
In this case, their fate depends on the relative orientation of their respective Burgers vectors, ${\vec{b} = \pm [0001]}$.  
If the two Burgers vectors point in the same direction, a double screw dislocation 
with its characteristic V-shaped step structure will emerge at the sample surface, 
cf.\ Fig.\,\ref{Fig:layer_thickness}(e,f) for a 40\,AL film.  
In contrast, if the two Burgers vectors have opposite signs, 
the partial screw dislocations will annihilate and form an extended smoothed step edge.
This appearance remains qualitatively unchanged up to the thickest Gd films studied here, i.e., 600\,AL.

\subsection{Electronic Properties}
\label{subsec:Elec}
\vspace{-0.3cm}
\begin{figure*}[htbp]
	\centering
	\includegraphics[width=\linewidth]{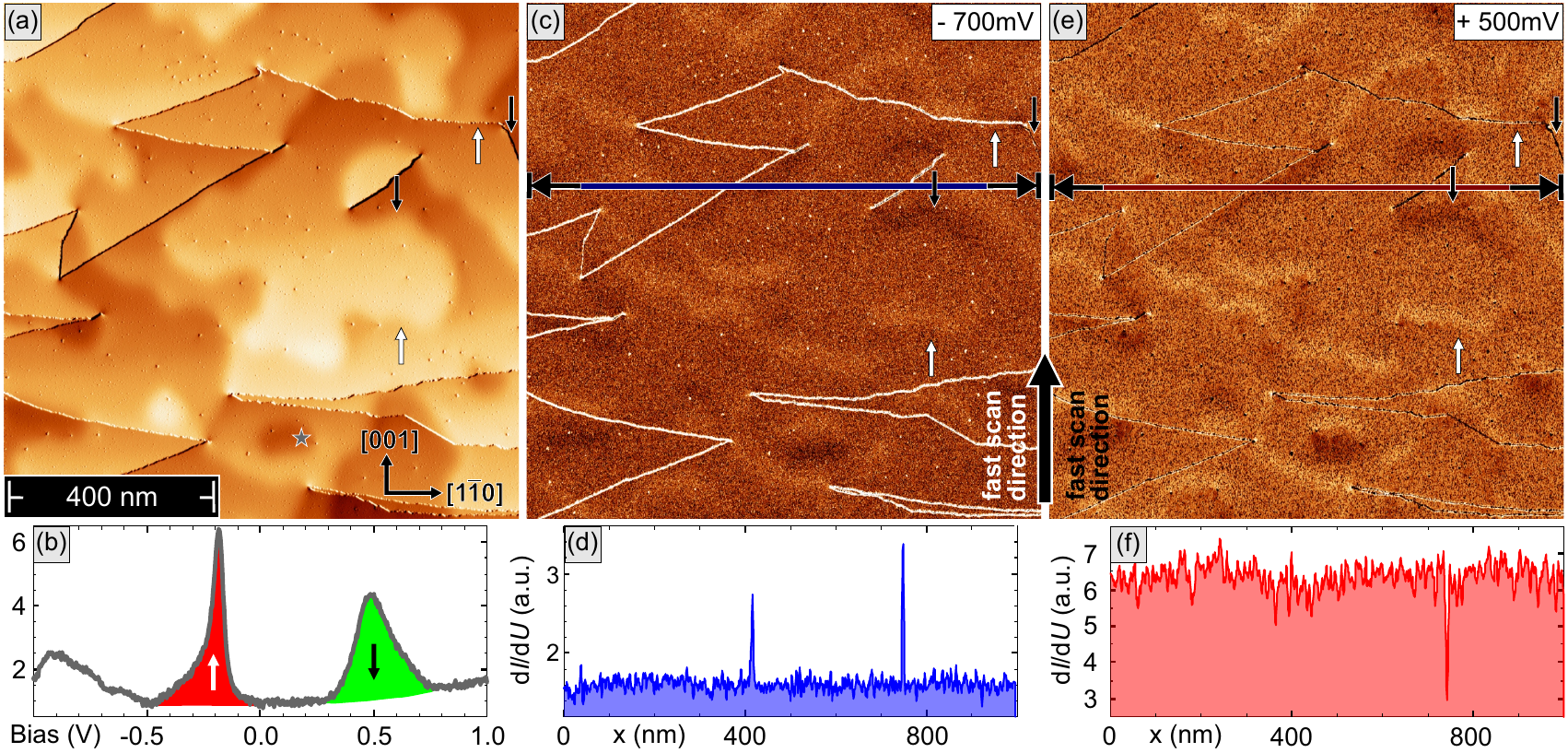}
			\caption{Spin-averaged STM measurements performed 
			on a 200\,AL thick Gd(0001) film on W(110) with an unpolarized W tip. 
			(a) The topography image shows a surface with few defects and the characteristic V-shaped double screw dislocations.  
			(b) Tunneling d$I$/d$U$ spectrum measured at the position marked by a grey star in (a). 
			The occupied majority (red) and the unoccupied minority (green) part of the $5d_{z^2}$-like surface state 
			can clearly be recognized as peaks.  
			(c) d$I$/d$U$ map of the same area shown in (a) measured at a bias voltage $U_{\text{bias}} = - 700$\,mV, 
			giving access to the density of states at the energy of the unoccupied part of the Gd(0001) surface state.  
			(d) Smoothed line profile of the d$I$/d$U$ signal taken along the blue-transparent line in (c).  
			(e,f) Same as (c,d), but now measured at $U_{\text{bias}} = + 500$\,mV.
			Stabilization current: $I_{\text{set}} = 1$\,nA.}
		\label{Fig:unpolarized}
\end{figure*}
Figure~\ref{Fig:unpolarized} summarizes spin-averaged STM measurements 
for a 200\,AL thick Gd(0001)/W(110) film taken with an unpolarized W tip (see supporting information in Ref.\,\onlinecite{SupplMat}). 
The film was prepared by room temperature deposition 
and post-growth annealing at ${T_{\rm anneal}\,\geq\,900\,\text{K}}$ for 5\,min. 
In agreement with the data presented so far, the STM topography image shown in Fig.\,\ref{Fig:unpolarized}(a) 
exhibits a surface morphology with characteristic double-screw dislocations and smoothed step edges.  
At the location marked with a grey star in Fig.\,\ref{Fig:unpolarized}(a), 
a d$I$/d$U$ point spectrum was taken which is plotted in Fig.\,\ref{Fig:unpolarized}(b).
It shows two pronounced peaks which originate from the well-known exchange-split $5d_{z^2}$-like surface states of Gd(0001). 
The occupied majority part appears at a binding energy of $eU = - 190$\,meV, 
whereas the empty minority part is observed at $eU = +480$\,meV, 
highlighted in red and green in Fig.\,\ref{Fig:unpolarized}(b), respectively. 
These values are in good agreement with the binding energies reported 
in earlier STS studies \cite{Bode1998,PhysRevLett.83.3017,Wegner2006}
and in a combined angle-resolved photoemission (PE) and inverse PE study by Weschke \emph{et al} \cite{Weschke1996}.

Figure \ref{Fig:unpolarized}(c,e) shows d$I$/d$U$ maps which were simultaneously obtained 
at bias voltages (c) $U_{\text{bias}} = -700$\,mV and (e) $U_{\text{bias}} = +500$\,mV. 
As evidenced by the respective slightly smoothed line profiles, presented in Fig.\,\ref{Fig:unpolarized}(d) and (f) below, 
in both cases the d$I$/d$U$ signal strength shows a relatively small variation, 
indicating an almost constant density of states (DOS) if measured with a non-magnetic probe tip. 
The only exceptions are a weak modulation which occurs on length scales of about 100\,nm and more narrow dips/peaks.  
Correlation with the topography image in Fig.\,\ref{Fig:unpolarized}(a) reveals that these variations 
of the d$I$/d$U$ signal are not related to the local electronic properties of the Gd film, 
but instead caused by the finite response time of the feedback circuit.  
A thick arrow between panels (c) and (e) indicates the fast scan direction of these data sets. 
Obviously, when scanning the tip over smoothed step edges or step edges, 
such as those marked by arrows in Fig.\,\ref{Fig:unpolarized}(a), 
the tip needs to be retracted whenever the sample height increases (upward slope, white arrows) 
and extended if the sample height decreases (downward slope, black arrows). 
Due to the finite response time of the feedback circuit this results in a subtle increase or decrease of the tunneling current $I$, respectively.
Comparison of the topography in Fig.\,\ref{Fig:unpolarized}(a) with the d$I$/d$U$ maps presented in panels (c) and (e) reveals 
that this variation of the $I$ results in a d$I$/d$U$ signal which is enhanced at every upward slope and reduced at downward slopes.  
The backward scan (opposite fast scan direction, not shown here) exhibits the opposite trend.
Therefore, we can conclude that the spin-averaged density of states of the surface of Gd films grown on W(110) 
as measured with non-magnetic probe tips is essentially homogeneous. 
As we will see below, this changes if magnetic probe tips are used and enables the imaging of magnetic domains.  

\vspace{-0.3cm}
\begin{figure*}[htbp]
	\centering
	\includegraphics[width=\textwidth]{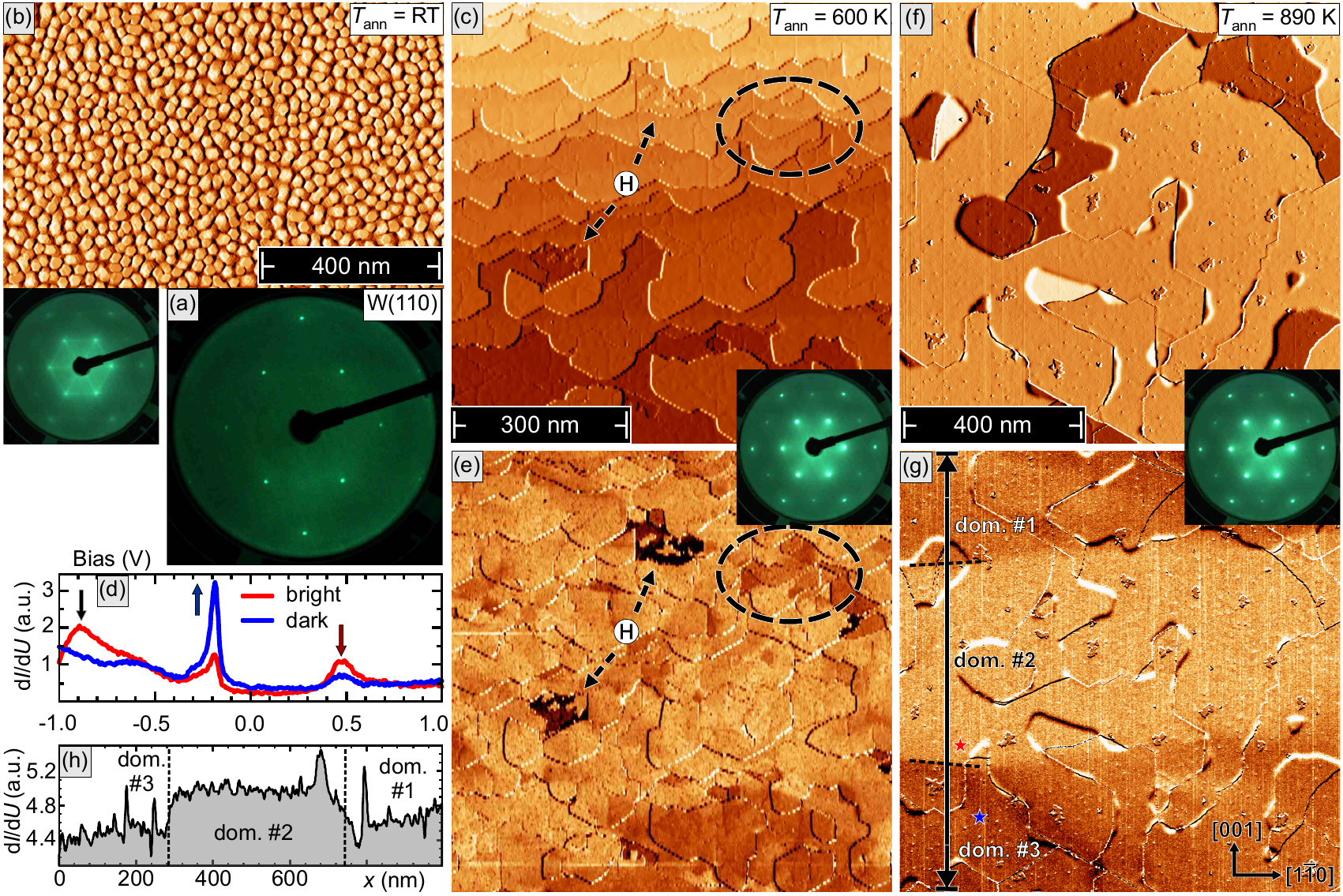}
			\caption{(a) LEED diffraction figure of the clean W(110) crystal. 
			(b) Overview scan of a 30\,AL thick Gd(0001) film grown on W(110) at room temperature without post-annealing.
			(c) Topography of a similar film as (b), but post-annealed at $T_{\text{ann}} = 600$\,K.
			(d) Typical d$I$/d$U$ point spectra measured with a magnetically sensitive Gd/W tip.  
			(e) Magnetically sensitive d$I$/d$U$ signal of the same sample surface presented in (c).
			Note that the dark contrast is not of magnetic origin but caused by the local adsorption of hydrogen (marked H). 
			(f) Topography and (g) magnetically sensitive d$I$/d$U$ signal of 30\,AL Gd(0001)/W(110) 
			upon annealing at $T_{\text{ann}} = 890$\,K. 
			(h) Line section of the d$I$/d$U$ signal measured along the arrows in (g).
			The insets show the evolution of the LEED pattern with increasing annealing temperatures. 
			LEED parameters: $T_{\text{LEED}} = 110$\,K, $E_{\text{LEED}} = 133$\,eV. 
			STM parameters: $U_{\text{bias}} = -700$\,mV, $I_{\text{set}} = 1$\,nA.}
		\label{Fig:annealingtemp}
\end{figure*}

\subsection{Magnetic Properties}
\label{subsec:Mag}
\vspace{-0.3cm}
\noindent
{\it Post-growth annealing temperature} --- It has been shown in numerous reports that the annealing temperature 
significantly affects the magnetic properties of Gd(0001) films grown on W(110) substrates 
\cite{PhysRevB.45.503,PhysRevB.50.6457,PhysRevB.52.1078,Tober1996}.
For example, 50\,nm thick films showed a reduced low-temperature coercivity if annealed at ${T_{\text{ann}} = 570}$\,K, 
an effect which was attributed to the annealing of defects \cite{PhysRevB.52.1078}.
It was argued that ``the domain wall mobility is enhanced with improving crystal quality'', 
thereby resulting ``in a reduction of the coercive field'' \cite{PhysRevB.52.1078}.
However, it has also been shown that too high annealing temperatures 
may result in discontinuous films \cite{PhysRevB.45.503,FLB1993,FL1994,Tober1996}.

In order to appropriately evaluate the effects the annealing temperature has on both 
the surface morphology and the magnetic domain structure, we exemplarily performed 
a combined LEED and SP-STM study of 30\,AL Gd/W(110), see Fig.\,\ref{Fig:annealingtemp}. 
Fig.\,\ref{Fig:annealingtemp}(a) displays the LEED pattern of clean W(110).  
As shown in the STM image of Fig.\,\ref{Fig:annealingtemp}(b), Gd deposition without any further annealing 
results in a very rough surface with numerous hexagon-shaped islands 
with a typical diameter of 20\,nm and several nanometer deep trenches. 
Similar to earlier reports \cite{PhysRevB.45.503}, the corresponding LEED pattern, which is shown 
as an inset in Fig.\,\ref{Fig:annealingtemp}(b), exhibits diffuse refraction spots, 
most likely resulting from poor long-range order. 

Figure~\ref{Fig:annealingtemp}(c) shows an STM image of a Gd film 
which was post-annealed at $T_{\text{ann}} = 600$\,K for 15 minutes. 
The surface topography is comparable to the data presented in Fig.\,\ref{Fig:layer_thickness}(a) further above.  
We recognize a decent terrace-and-step growth without any screw dislocations.  
However, the terraces are not flat but exhibit a large number of line defects, 
as already discussed in Sec.\,\ref{subsec:Struc}, 
indicating a high density of grain boundaries between differently stacked patches of the film. 
The LEED pattern displayed in the inset shows sharp refraction spots, thereby confirming a much improved crystallinity 
as compared to the as-grown film presented in Fig.\,\ref{Fig:annealingtemp}(b). 

Typical d$I$/d$U$ spectra obtained with a spin-polarized Gd/W tip 
on two locations of a Gd/W(110) film are shown in Fig.\,\ref{Fig:annealingtemp}(d).  
The locations were chosen somewhat arbitrarily such that the spectra show maximum contrast.  
We recognize two spectra which are very similar qualitatively, i.e., if solely judged 
on the basis of the general spectral shape and the peak positions, 
but which---at least at some bias voltages---differ quite significantly in terms of the d$I$/d$U$ signal strength.  
In close accordance with the spin polarization of Gd(0001) 
determined by spin-resolved inverse photoemission spectroscopy \cite{Donath1996} 
and previously published SP-STS results obtained with Fe-coated probe tips \cite{Bode1998,APL1999}, 
we find a contrast reversal between the occupied majority [marked $\uparrow$ in Fig.\,\ref{Fig:annealingtemp}(d)] 
and the unoccupied minority part ($\downarrow$) which takes place at a bias voltage $U_{\rm bias} \approx +350$\,mV.  
However, with the Gd tips employed here, we find an additional contrast reversal 
in the occupied states at $U_{\rm bias} \approx -500$\,mV. 
It appears to be related to a peak which is located at around $U \approx -900$\,mV, 
see black arrow in Fig.\,\ref{Fig:annealingtemp}(d).  
We tentatively attribute this feature in our STS data to a bundle of Gd bands 
which originate from hybridizing majority $5d$- and $6s$-derived states 
and disperse in the $\overline{\Gamma {\rm K}}$ direction of the surface Brillouin zone \cite{Kurz2002}. 
As described in Figs.\,8 and 9 of Ref.\,\onlinecite{Kurz2002}, these bands can be found 
in the energy range between 0.5\,eV and 1.0\,eV below the Fermi level, exhibit a very flat dispersion 
and, therefore, are expected to result in an appreciable majority DOS in the surface layer. 
We would like to note that the bias voltage range just below this peak, especially around $U \approx -700$\,mV,
turned out to be particularly suited for spin-resolved d$I$/d$U$ mapping experiments with Gd/W tips, 
as the magnetic contrast was most reliable and rather stable.  

The spin-dependent contrasts identified in Fig.\,\ref{Fig:annealingtemp}(d) can be utilized 
to map the magnetic domain structure of the Gd(0001) surface. 
Fig\,\ref{Fig:annealingtemp}(e) presents a spin-resolved d$I$/d$U$ map 
which was obtained with a magnetically sensitive Gd/W tip 
simultaneously with the topographic image of Fig.\,\ref{Fig:annealingtemp}(c).
The contrast variation here is caused by spin-polarized tunneling and reflects the domain structure of the film.\footnote{
	Note that the much darker patches (marked by arrows) are caused by the local adsorption of hydrogen \cite{Getzlaff1999}.  
	This signal does not contain any magnetic information but only indicates the quenching of the Gd surface state.  }
The magnetic domains visible in Fig.\,\ref{Fig:annealingtemp}(e) are quite irregular,
e.g.\ in the region marked by hatched black ellipses, and often form tiny patches. 
We speculate that this domain structure is the result of the pinning of magnetic domain walls 
at structural defects, such as step edges or structural grain boundaries. 

As evidenced by the data presented in Fig.\,\ref{Fig:annealingtemp}(f) and (g), 
annealing of the films profoundly changes the magnetic domain structure of Gd films on W(110).  
Again, a 30\,AL thick Gd film was deposited at room temperature, but the post-annealing temperature 
is increased to $T_{\text{ann}} \approx 890$\,K for 15\,min. 
In comparison to Fig.\,\ref{Fig:annealingtemp}(c), the topographic STM image 
shown in Fig.\,\ref{Fig:annealingtemp}(f) reveals a much improved surface quality
with a strongly reduced density of step edges and structural grain boundaries.  
At the same time, the LEED pattern (inset) is basically undistinguishable from the one 
presented in Fig.\,\ref{Fig:annealingtemp}(c), indicating that the short-range crystalline order 
on length scales of the coherence length of low-energy electrons ($\approx 100$\,nm) remains unchanged.  
Although this annealing results in the formation of double-screw dislocations, 
as described before in Fig.\,\ref{Fig:layer_thickness}, the magnetic domain structure changes considerably, 
as can be seen in the spin-sensitive d$I$/d$U$ map shown in Fig.\,\ref{Fig:annealingtemp}(g).   
On a $1\,\mu{\rm m} \times 1\,\mu{\rm m}$ scan range we recognize only three magnetic domains, 
 marked as dom.\,\#1 through dom.\,\#3, which are separated by domain walls 
(highlighted by hatched lines) roughly oriented along the W$[1\bar{1}0]$ direction.  
A line profile of the d$I$/d$U$ signal along the arrow across the three domains is plotted in Fig.\,\ref{Fig:annealingtemp}(h). 
In contrast to the essentially flat data taken with non-magnetic tips, cf.\ Fig.\,\ref{Fig:unpolarized}(d) and (f), 
the line profile presented here features two contrast levels.  
While a lower d$I$/d$U$ signal is recorded at domains \#1 and \#3, a higher signal is measured for domain \#2.
Obviously, the reduced density and different kind of structural defects upon annealing to 890\,K
leads to a strong reduction of the pinning potential for magnetic domain walls.  
As a result, the film is able to adapt a more uniform domain structure.

Although we are not able to unambiguously determine whether the magnetic domain structures 
imaged in Fig.\,\ref{Fig:annealingtemp}(g) represents an in-plane or an out-of-plane contrast, 
we interpret the strongly anisotropic course of the domains walls along the W$[1\bar{1}0]$ direction 
as evidence for a magnetization which is oriented in-plane along this direction.  
Thereby, the magneto-static energy associated with head-to-head or tail-to-tail spin configurations can be avoided.
Unfortunately, a comparison of these data with previous experimental results 
is in many cases complicated by an insufficient documentation of the scattering geometry.  
One of the few exceptions is the study of Weller {\em et al.}, \cite{Weller1986} 
where it is explicitly stated that their spin-polarized low-energy electron diffraction (SPLEED) experiments 
are essentially sensitive to a magnetization along the W$[1\bar{1}0]$ direction, in agreement with our hypothesis above.

\begin{figure*}[t]
	\includegraphics[height=0.8\textheight]{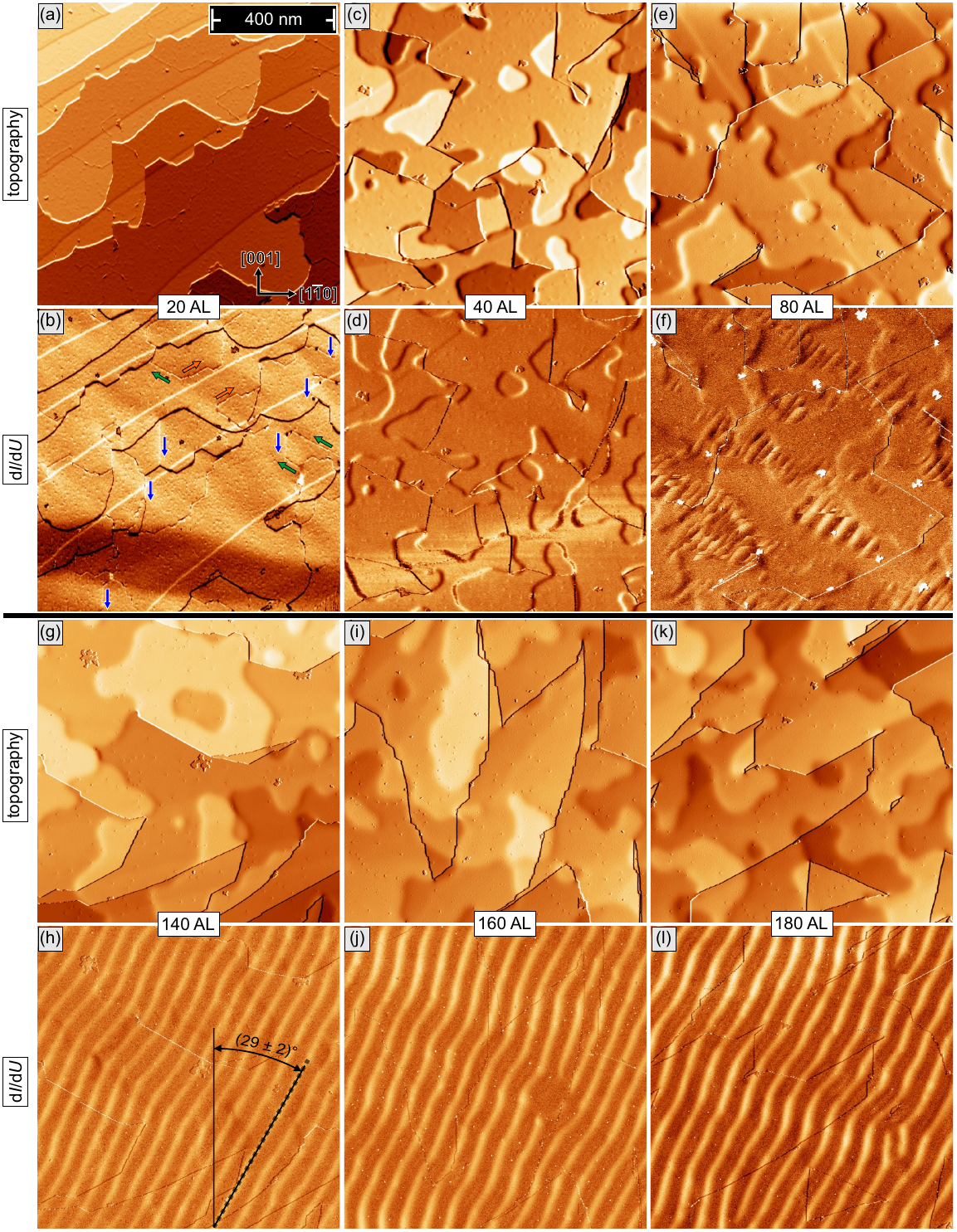}
	\caption{First part of a coverage-dependent series of topographic images (top) and the simultaneously measured 
    		magnetically sensitive d$I$/d$U$ maps (bottom) of Gd(0001) films grown on a W(110) substrate
		(continued in Fig.\,\ref{Fig:thickness_p2}). 
		While a few large domains with domain walls running along the W$[1\bar{1}0]$ direction 
		can be recognized at low coverage, i.e., for (a,b) 20\,AL and (c,d) 40\,AL, stripe domains which are tilted 
		by an angle $\alpha = (25 ... 30)^{\circ}$ against the W$[001]$ direction can be found between (g,h) 140\,AL and (k,l) 180\,AL.	
		A transitional state between the two regimes is shown in (e,f).
		Scan parameters: $U_{\text{bias}} = -700$\,mV, $I_{\text{set}} = 1$\,nA.}
    \label{Fig:thickness_p1}
\end{figure*}
\begin{figure*}[t]
	\includegraphics[height=0.8\textheight]{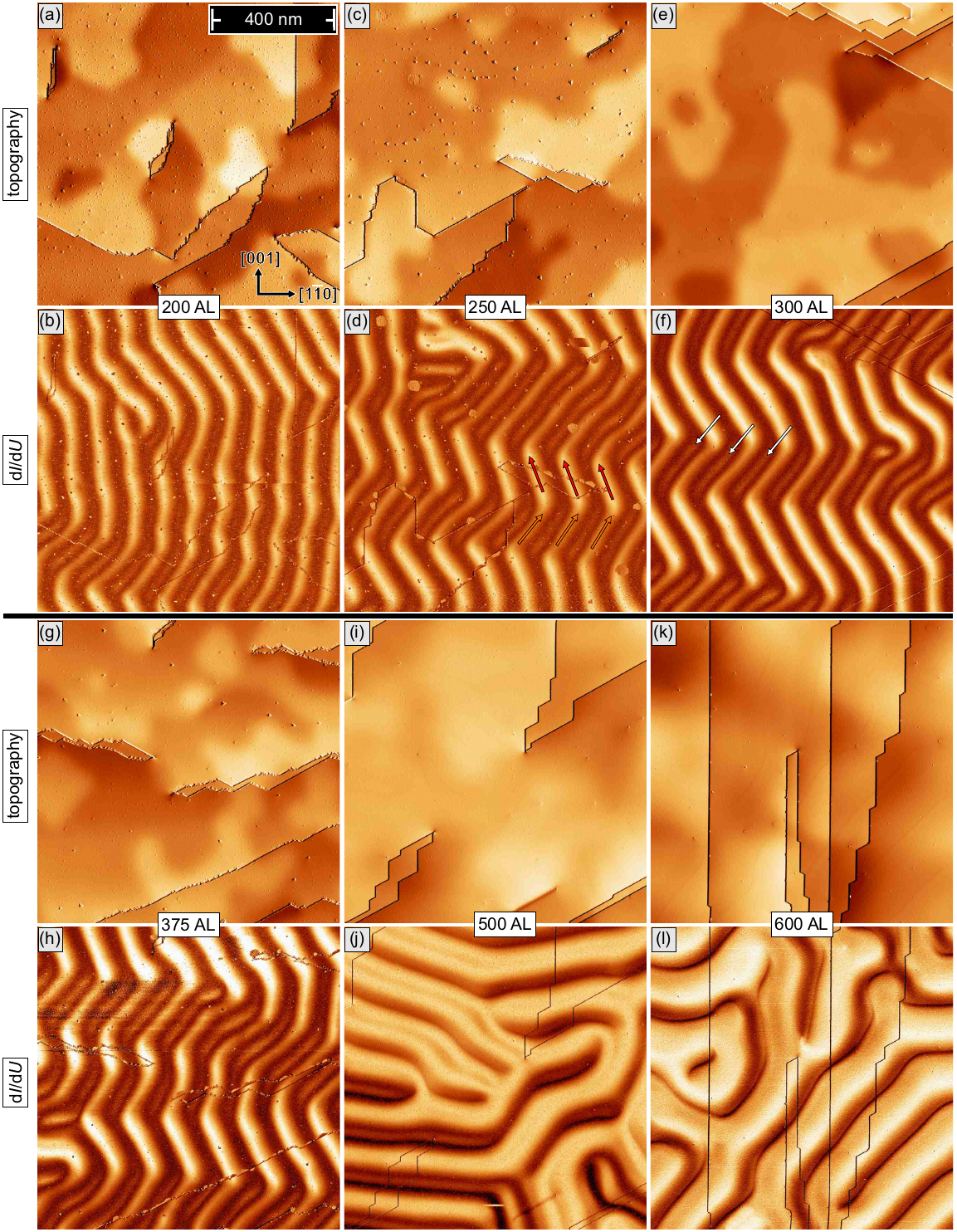}
	\caption{Second part of a coverage-dependent series of Gd(0001) films grown on a W(110) substrate 
		(continued from Fig.\,\ref{Fig:thickness_p1}). 
		At coverages between (a,b) 200\,AL and (g,h) 375\,AL the orientation of these stripe domains 
		periodically switches between positive and negative tilt angles, resulting in an apparent zig-zag like domain pattern. 
		For even thicker films between (i,j) 500\,AL and (k,l) 600\,AL a worm-like magnetic domain patterns 
		with even larger periodicities is found. 
		Scan parameters: $U_{\text{bias}} = -700$\,mV and $I_{\text{set}} = 1$\,nA.
		}
    \label{Fig:thickness_p2}
\end{figure*}
{\it Film thickness-dependent domain structure} --- We have performed an extensive SP-STM investigation 
of the thickness-dependent magnetic domain structures of Gd films grown on W(110). 
As described before, the films were deposited at room temperature and subsequently annealed 
at $T_{\rm anneal} \geq 900$\,K for five minutes on an electron-beam heating stage. 
We would like to note that a large part of the data presented in Figs.\,\ref{Fig:thickness_p1} and \ref{Fig:thickness_p2} 
have been obtained in a relatively short experimental run 
with the identical tip for which we can exclude even minor tip changes, see Ref.\,\onlinecite{Note3}. 
The complete set of data consists of 12 different coverages. 
The first, low-coverage part of this study is presented in Fig.\,\ref{Fig:thickness_p1}. 
The six data sets displayed are arranged in two rows divided by a horizontal black line. 
Each row is subdivided into three columns which show the STM topography in the top 
and the corresponding, simultaneously measured magnetic d$I$/d$U$ map in the bottom image. 
All images show a scanned area of $1\,\mu{\rm m} \times 1\,\mu{\rm m}$. 

Starting at 20\,AL, the Gd film thickness is still below the onset 
where relaxation processes set in, cf.\ Fig.\,\ref{Fig:layer_thickness}(h). 
Correspondingly, the topographic image in Fig.\,\ref{Fig:thickness_p1}(a)
 exhibits flat terraces separated by single-atomic step edges.  
On the terraces numerous trenches can be recognized which are characteristic 
for the structural grain boundaries between differently stacked patches.
Yet, as can be seen in the magnetically sensitive d$I$/d$U$ map of Fig.\,\ref{Fig:thickness_p1}(b), 
due to the high annealing temperature their density is sufficiently reduced to allow for large domains. 
Domain walls are roughly oriented along the W$[1\bar{1}0]$ direction.  
Within individual domains a weak spatial variation of the d$I$/d$U$ signal can be recognized. 
For example, the bright domain which covers more than the upper half of the image, 
exhibits areas with a slightly enhanced or reduced d$I$/d$U$ signal. 
Comparison of the structural and magnetic data displayed in Fig.\,\ref{Fig:thickness_p1}(a) and (b), respectively, 
reveals that the modification of the magnetic d$I$/d$U$ signal correlates with the orientation 
of the structural grain boundaries which separate differently stacked patches of the film, cf.\ Fig.\,\ref{Fig:layer_thickness}(c).
Due to the hexagonal symmetry of the Gd surface layer, the structural grain boundaries come in three orientations, 
marked by blue, red, and green arrows in Fig.\,\ref{Fig:thickness_p1}(b).  
We speculate that the grain boundaries are accompanied by an additional in-plane anisotropy term 
which causes the magnetization to locally deviate from the W$[1\bar{1}0]$ direction. 
This deviation leads to an improved or poorer alignment with the tip magnetization, 
resulting in an enhanced or reduced d$I$/d$U$ signal due to the TMR effect, respectively.  
In Fig.\,\,\ref{Fig:thickness_p1}(b) regions with grain boundaries
oriented along ${\text{Gd}[1\bar{1}00] \parallel \text{W}[001]}$ (blue arrows) appear brighter, 
whereas a darker contrast is observed in regions where they are oriented 
along $\text{Gd}[\bar{1}010]$ (green) or $\text{Gd}[01\bar{1}0]$ (red).  
It appears that these dislocation-induced deviations of the magnetization 
are also responsible for some if not all fluctuations of the domain wall orientation.

Inspection of the STM image of a 40\,AL film, displayed in Fig.\,\ref{Fig:thickness_p1}(c), 
reveals that the structure of the film has changed considerably.  
The presence of numerous double-screw dislocations and smooth step edges evidences 
that the film thickness is above the threshold for the relaxation, cf.\ Fig.\,\ref{Fig:layer_thickness}(i). 
Nevertheless, the domain structure represented in Fig.\,\ref{Fig:thickness_p1}(d) 
is essentially the same as for the 20\,AL film of Fig.\,\ref{Fig:thickness_p1}(a,b), 
as it still shows a few extended domains with domain walls along the W$[1\bar{1}0]$ direction.

When increasing the Gd coverage further to 80\,AL up to 180\,AL the structural properties 
of the surface as imaged by STM remain largely unchanged, see Fig.\,\ref{Fig:thickness_p1}(e,g,i,k).  
In either case the topography of the surface is dominated by double-screw dislocations, 
smoothed step edges, and numerous point-like defects indicating the presence of edge dislocations.  
Yet, the magnetic domain structure changes considerably.  
For a Gd coverage of 80\,AL, Fig.\,\ref{Fig:thickness_p1}(f), 
we observe several triangle- or lens-shaped areas which exhibit a stripe pattern, 
whereas the surrounding surface still shows a homogeneous d$I$/d$U$ signal.  
The lateral size of the striped areas typically amounts to a few hundred nanometers.  
When performing several experimental runs we found this inhomogeneous surface magnetic structure 
at nominal coverages of $(100 \pm 20)$\,AL.   
We speculate that the variation originates from slight fluctuations of the growth parameters which are beyond our control. 

At even higher coverage, between 140\,AL and 180\,AL, Fig.\,\ref{Fig:thickness_p1}(h,j,l), 
pronounced periodic stripes cover the entire sample surface.  
As exemplarily indicated in Fig.\,\ref{Fig:thickness_p1}(h), these stripes are initially tilted 
by about $(29 \pm 2)^{\circ}$ with respect to the $[001]$ direction of the W substrate.  
We would like to note that negative tilt angles also occurred, 
as exemplarily shown in Fig.\,S4(b) of the Ref.\,\onlinecite{SupplMat}.
Already a superficial inspection suggests that the periodicity of these stripes increases with coverage. 
Furthermore, an increasing tendency towards orientational deviations from straight stripes becomes manifest.  
The data presented in Fig.\,\ref{Fig:thickness_p1}(j,l) suggest that the dark stripes are broader than the bright stripes. 
Close inspection reveals that for these 160\,AL and 180\,AL thick films a faint intermediate intensity is visible within the dark regions. 
As will be pointed out below these effects become more pronounced for even thicker films.

\begin{figure}[t]
	\centering
	\includegraphics[width=\columnwidth]{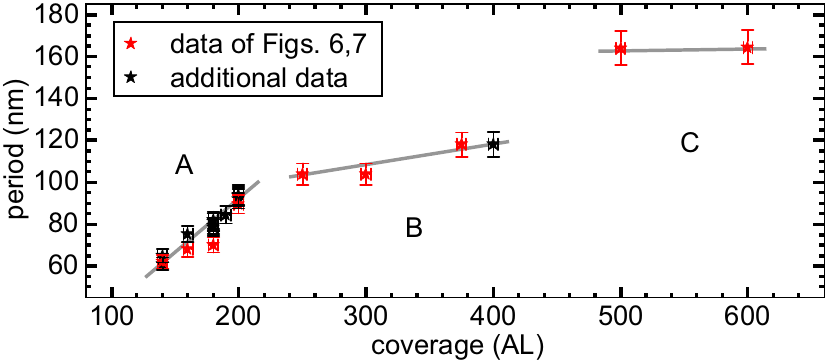}
			\caption{Periodicity of the stripe domains as extracted from the data of Figs.\,\ref{Fig:thickness_p1} 
			and \ref{Fig:thickness_p2} (red) and other SP-STM data not presented here (black).
			Grey lines serve as guides to the eye only.  Three thickness ranges can be distinguished:  
			(A)\,Between 140\,AL and 200\,AL the periodicity increase linearly with Gd film thickness. 
			(B) The periodicity increase levels off between 200\,AL\,$\le \Theta \le 400$\,AL. 
			Another increase occurs between 400\,AL\,$< \Theta < 500$\,AL (C). }
		\label{Fig:thickness_plot}
\end{figure}

These impressions are further substantiated by the data taken on Gd films 
with higher coverages which are presented in Fig.\,\ref{Fig:thickness_p2}.
For Gd coverages between 200\,AL, Fig.\,\ref{Fig:thickness_p2}(a,b), up to 375\,AL, Fig.\,\ref{Fig:thickness_p2}(g,h) 
the stripe orientation more or less periodically switches between positive and negative tilt angles, 
resulting in an apparent zig-zag like domain pattern. 
Furthermore, the brightness of the stripes which are tilted by positive and negative angles, becomes increasingly different.  
These data clearly reveal that the magnetic d$I$/d$U$ signal for some stripes 
no longer oscillates between a minimal and maximal value but exhibits intermediate maxima.
Faint bright lines, three of which are marked by white arrows in Fig.\,\ref{Fig:thickness_p2}(f),
become visible within the dark segments of the stripes which are rotated in the clockwise direction.  
These observations will be discussed in the following Sec.\,\ref{sec:Discussion} in detail.

Finally, for very thick films with a Gd coverage of 500\,AL, Fig.\,\ref{Fig:thickness_p2}(i), 
or even 600\,AL, Fig.\,\ref{Fig:thickness_p2}(k), the surface topography becomes more and more smooth 
with fewer and fewer double-screw dislocations, step edges, and edge dislocations.  
This change in film structure is accompanied by a magnetic domain structure 
where the domains and domain walls no longer follow certain crystallographic directions, 
but instead wind quite irregular around one another, see Fig.\,\ref{Fig:thickness_p2}(j,k).  
Nevertheless, in close similarity to the dark segments observed for the clockwise-rotated stripes 
in the magnetically sensitive d$I$/d$U$ maps of Fig.\,\ref{Fig:thickness_p2}(b,d,f,h), 
the stripe pattern displays intermediate minima and maxima of the d$I$/d$U$ signal 
in the bright and dark stripes, respectively.  

Figure~\ref{Fig:thickness_plot} summarizes the periodicities of the stripe domains extracted 
from Figs.\,\ref{Fig:thickness_p1} and \ref{Fig:thickness_p2} for Gd film thicknesses between 140\,AL and 500\,AL. 
We can recognize three coverage ranges, labelled A--C in Fig.\,\ref{Fig:thickness_plot}.
A linear increase is obtained in A, where the periodicity increases from $(62 \pm 3)$\,nm 
at a 140\,AL up to $(91 \pm 5)$\,nm for a 200\,AL thick Gd film.
For even thicker Gd films, i.e.\ in the thickness range B between 200\,AL\,$ \le \Theta \le 400$\,AL, 
the periodicity increase seems to level off until another increase occurs towards C between 400\,AL\,$< \Theta < 500$\,AL. 

\section{Discussion}
\label{sec:Discussion}
\vspace{-0.3cm}
The data presented in Figs.\,\ref{Fig:thickness_p1} and \ref{Fig:thickness_p2} can be explained qualitatively 
by the subtle interplay between the exchange energy, the magneto-static energy related to the stray field, 
and magneto-crystalline contributions to the anisotropy. 
For Gd/W(110) these magnetic properties are---in comparison to bulk Gd---strongly 
modified by the presence of substantial epitaxial strain. 
For thin films up to about 60\,AL we find rather large domains, cf.\ Figs.\,\ref{Fig:annealingtemp}(g) and \ref{Fig:thickness_p1}(b,d).  
SP-STM images with a scan range of $1\,\mu{\rm m} \times 1\,\mu{\rm m}$ typically show 1-2 domain walls only,  
roughly oriented along the W$[1\bar{1}0]$ direction.   

\begin{figure*}[t]   
			\includegraphics[width=1\linewidth]{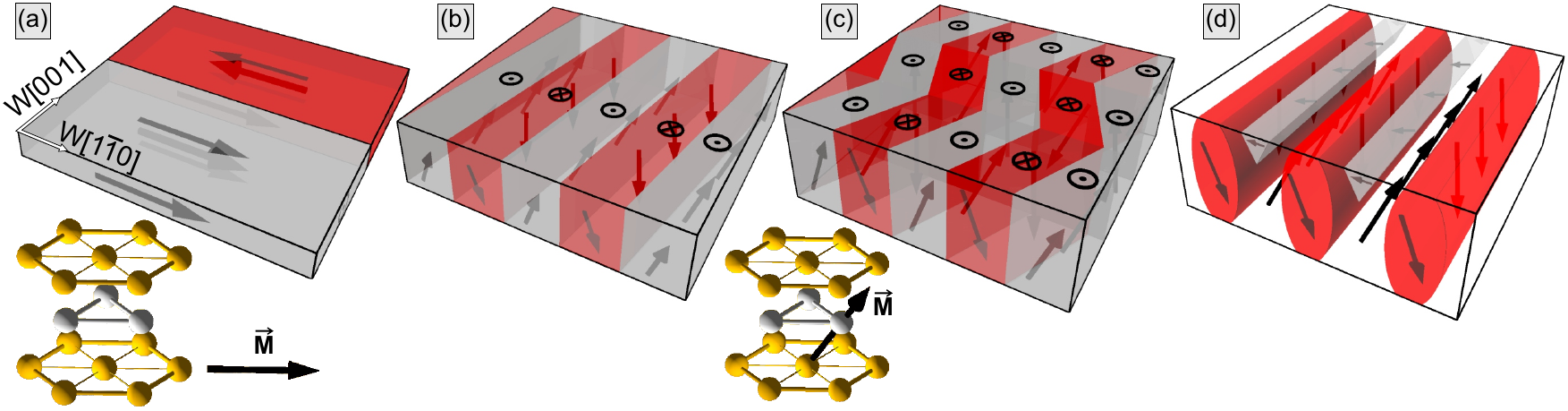}%
				\caption{Schematic representation of the thickness-dependent magnetic domain structure of Gd(0001) films grown on W(110). 
			(a) For thin films, $\Theta > \Theta_{\text{crit}} \approx (100 \pm 20)$\,AL, we find large domains 
			with an \textit{in-plane} easy axis oriented along ${\text{Gd}[11\bar{2}0] \parallel \text{W}[1\bar{1}0]}$, 
			as indicated by a black arrow in the model of the hcp crystal structure. 
			(b) Just above $\Theta_{\text{crit}}$ we find tilted stripe domains,  
			consistent with a magnetization along the Gd$[20\bar{2}3]$ direction. 
			(c) Around 300\,AL the stripe domains form a zig-zag pattern.
			(d) At $\Theta \geq 500$\,AL branching domains form.}
		 \label{Fig:magnetizationmodel}
	\end{figure*}

As sketched in Fig.\,\ref{Fig:magnetizationmodel}(a), this observation is consistent with films magnetized along this in-plane axis. 
It appears that---even though the easy axis of bulk Gd is tilted by $30^{\circ}$ from the c-axis---a 
perpendicular or canted magnetization relative to the sample plane is energetically unfavorable for thin films. 
This could either be caused by (i) the relatively large magneto-static energy associated with 
any significant out-of-plane magnetization, or (ii) by the uniaxial strain of the hexagonal Gd(0001) lattice 
due to epitaxial growth on the rectangular W(110) unit cell (see Fig.\,\ref{Fig:crystalmodels}) 
which might modify the magneto-crystalline anisotropy energy density. 
Also a combination of the two effects cannot be excluded.    

As the film thickness exceeds a critical value $\Theta_{\text{crit}} \approx (100 \pm 20)$\,AL, 
a spin reorientation transition (SRT) from in-plane to out-of-plane is observed.  
While we are unable to identify the exact physical origin of the SRT in Gd/W(110), 
it is obvious that, at this coverage, it becomes energetically unfavorable to align the magnetization with the film plane.  
If the magneto-static energy was responsible for the in-plane magnetization [scenario (i)], 
the energy penalty related to the film's magneto-crystalline anisotropy 
overcompensates this effect for coverages $\Theta > \Theta_{\text{crit}}$. 
Likewise (ii), it might be possible that the epitaxial strain relaxes to an extent 
unsuitable to maintain a magnetization along the in-plane axis.  

Upon rotation of the easy axis to out-of-plane, the Gd film forms stripe domains 
which result in a flux-closure configuration between adjacent domains and thereby reduce the sample's stray field.  
Stripe domains in perpendicularly magnetized thin films have frequently been observed and intensively been discussed in the literature. 
They originate from the competition between exchange interaction which favors a parallel orientation of the spins, 
and the magneto-static energy associated to the stray field which can be minimized by the formation of antiparallel domains.

Particular attention was paid to the properties of stripe domains in the vicinity of a spin reorientation transition (SRT)
where the easy axis of magnetization rotates from out-of-plane to in-plane or {\em vice versa}.   
It has generally been observed on $3d$ transition metal films with a thickness 
of a few atomic layers only \cite{Speckmann1995,Oepen1997,PhysRevLett.93.117205}, 
that the periodicity of the out-of-plane magnetized stripe domains 
decreases as one approaches the critical thickness $\Theta_{\text{crit}}$ of the SRT. 
Our results presented in Fig.\,\ref{Fig:thickness_plot} are in qualitative agreement with these results. 
However, in ultra-thin film systems---irrespective whether the SRT occurs 
from out-of-plane at low film thickness to in-plane for thick films 
or {\em vice versa}---the stripe density usually increases {\em exponentially} 
as one approaches the SRT \cite{PhysRevLett.93.117205,Won2005}.
In contrast, for Gd films on W(110) we find a stripe periodicity which increases {\em linearly} with film thickness 
from just above $\Theta_{\text{crit}} \approx (100 \pm 20)$\,AL up to about 200\,AL, cf.\ Fig.\,\ref{Fig:thickness_plot}.

An analogous SRT with a critical film thickness of 27\,nm and a linearly increasing stripe width
has been found for FePd films \cite{Gehanno1999}.
The different behavior was assigned to the fact that the FePd films were much thicker than the ultra-thin films mentioned above.
While the spins in ultra-thin films can be considered as rigidly tightened together 
in the vertical direction by the exchange interaction, this is not necessarily the case for much thicker films.  
As a result, the magnetization component normal to the film plane, $m_z$, can vary with the distance from the interface \cite{Gehanno1999}. 
Although we are not able to present a conclusive picture of how the experimentally observed domain periodicities 
can be explained, we speculate that the major qualitative changes of the magnetic pattern 
from stripe domains to zig-zag stripe domains and branching domains might also be responsible 
for the reduced slope of the periodicity increase for Gd films with $\Theta > 200$\,AL 
as well as for the further increase at 400\,AL\,$< \Theta < 500$\,AL.

Interestingly, when reaching the critical film thickness $\Theta_{\text{crit}} = (100 \pm 20)$\,AL,
we could image indications for an inhomogeneous nucleation of stripe domains, cf.\ Fig.\,\ref{Fig:thickness_p1}(f). 
Although we cannot strictly exclude that the weak stray field of the Gd/W tip 
influences these nucleation processes \cite{AFM_ProbeTips,Kubetzka2003}, 
our data are in good qualitative agreement with earlier photoemission electron microscopy (PEEM) experiments, 
where the local melting of stripe domains was observed, thereby giving rise to paramagnetic patches \cite{Won2005}. 

At coverages $\Theta > \Theta_{\text{crit}}$, the data of Fig.\,\ref{Fig:thickness_p1}(h,j,l) and Fig.\,\ref{Fig:thickness_p2}(b,d,f,h) 
showed stripe domains which were not oriented along a low-index in-plane axis of the W(110) substrate,
but rotated within the film plane by about $\pm 30^{\circ}$ with respect to the W$[001]$ direction.  
At the same time we know from literature that the bulk easy magnetization direction 
of Gd is tilted by $30^{\circ}$ relative to the c-axis. 
Both conditions can be approximately fulfilled simultaneously by a magnetization which is, for example, 
oriented along the Gd$[20\bar{2}3]$ direction, indicated by a black arrow in the inset of Fig.\,\ref{Fig:magnetizationmodel}(b). 

This vector includes an angle of $\approx 36^{\circ}$ with respect to the c-axis 
and its projection onto the (0001) surface is tilted by $30^{\circ}$ to the W$[001]$ and the Gd$[\bar{1}100]$ direction.  
The resulting magnetic domain structure which is schematically represented in the sketch of Fig.\,\ref{Fig:magnetizationmodel}(b) 
is in very good agreement with the conclusions drawn by Berger and co-workers 
based on susceptibility measurements \cite{PhysRevB.50.6457,BergerJMMM,PhysRevB.52.1078}.  
Also the critical film thickness $\Theta_{\text{crit}} \approx (100 \pm 20)$\,AL identified by us 
agrees reasonably well with the value of 40\,nm determined by Berger and co-workers, 
which corresponds to 138\,AL \cite{BergerJMMM,PhysRevB.52.1078}.  

As the film thickness is increased beyond approximately 200\,AL, several interesting effects can be observed. 
Firstly, the stripes are no longer straight.  
Instead of extended areas with uniformly oriented domains, we find stripes 
the orientation of which frequently changes between $+30^{\circ}$ and $-30^{\circ}$, 
thereby forming a zig-zag pattern with a typical distance between turning points of several hundred nanometers. 
This zig-zag pattern may be related to magneto-elastic interactions 
caused by the uniaxial strain associated to the growth of Gd on a W(110) crystal surface. 
Since the crystallographic symmetry of the W(110) surface imposes two mirror lines, 
i.e., the $[001]$ and the $[1{\bar 1}0]$ axis, any situation where the domain and domain wall orientation 
uniformly deviate from these axes would result in a net magneto-elastic strain 
which is incompatible to the ``elastic environment{''}, see Sec.\,3.3.1 in Ref.\,\onlinecite{Hubert2008}.  
By forming magnetic domains which more or less periodically switch their orientation 
between $+30^{\circ}$ and $-30^{\circ}$, any global strain away from the above-mentioned axes is avoided.  
 
Secondly, the magnetic contrast observed for stripe domains which are rotated by $+30^{\circ}$ and $-30^{\circ}$ 
with respect to the W$[001]$ direction is strikingly different, cf.\ Fig.\,\ref{Fig:thickness_p2}(d,f,h).  
In these data sets---all measured with the same magnetic tip\footnote{%
	The identity of the tip apex was confirmed by the comparison of tunneling spectra 
	measured on the surfaces of Fig.\,\ref{Fig:thickness_p2}(d,f,h).  
	Since the energetic position of the $d_{z^2}$--like surface states remains unchanged 
	for Gd films with a local coverage $\Theta_{\rm loc} > 4$\,AL\cite{GETZLAFF1998}, 
	any modification of the recorded d$I$/d$U$ spectra could safely be attributed to a tip change.  
	However, the spectra measured on the sample surfaces of Fig.\,\ref{Fig:thickness_p2}(d,f,h) were virtually identical (not shown here), 
	thereby verifying the fact that the tip apex remained unchanged during these experiments.}%
---the (anti)-clockwise rotated domains appear much darker (brighter).  
Since purely out-of-plane domains would be degenerate with respect to the rotational sense, 
we conclude that a significant in-plane component also exists. 
We speculate, that this canted magnetization is the result of a significant magneto-crystalline contribution.
As discussed in the context of Fig.\,\ref{Fig:magnetizationmodel}(b), we expect 
that the in-plane component of the magnetization is oriented along the $[20\bar{2}3]$ or equivalent directions, 
such that the moments are along the easy axis of bulk Gd and at the same time aligned with the stripes domains. 
The schematic representation of the zig-zag domain structure is presented in Fig.\,\ref{Fig:magnetizationmodel}(c).

Thirdly, the magnetic d$I$/d$U$ signal of some domains of Gd films with a thickness
between 250\,AL and 375\,AL no longer oscillates between a minimal and maximal value, 
as observed for thinner films just above the SRT, but exhibits intermediate maxima, cf.\ Fig.\,\ref{Fig:thickness_p2}(d,f,h).  
This finding may be associated with the formation of branching domains, 
as originally proposed by Lifschitz \cite{Lifschitz1992}. 
The general concept of branching domains is that ``a progressive domain refinement towards the surface 
by iterated generations of domains'' results in a ``gain in closure energy'', see Sec.\,3.7.5 in Ref.\,\onlinecite{Hubert2008}.
We interpret the appearance of the intermediate maximum as the first set of branching domains.  
A potential domain structure with in-plane closure domains is sketched in Fig.\,\ref{Fig:magnetizationmodel}(d). 

We would like to note, however, that the intermediate maxima could initially only be observed 
in regions where the darker, clockwise rotated domains appear.  
No indication could be found in the anti-clockwise rotated stripe domains. 
We can only speculate about the reason for this asymmetry.  
One explanation would be that (i) branching domains exclusively exist for one stripe orientation, 
though we were not able to identify a simple explanation of how this directional selectivity might come about.  
Another explanation is (ii) related to the fact that the Gd/W tip we are using is known to exhibit a significant in-plane component. 
As we will show by data taken with different tips below, the branching domain increases 
the magnetic d$I$/d$U$ signal in some regions, but has no significant effect in another region.  

\begin{figure}[t]
	\centering
	\includegraphics[width=\columnwidth]{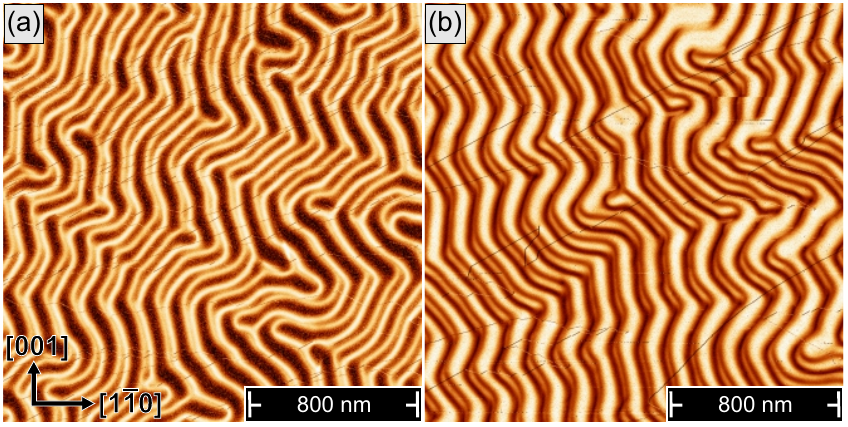}
			\caption{Two magnetically sensitive d$I$/d$U$ maps (scan size: $2\,\mu{\rm m} \times 2\,\mu{\rm m}$ each) 
			measured on the same 400\,AL Gd/W(110) with different Gd/W tips prepared by poking into the Gd film. 
			The data provide strong evidence that---depending on the tip's quantization 
			axis---some branching domains may remain undetected (see text for details). 
			Scan parameters: $U_{\text{bias}} = -700$\,mV and $I_{\text{set}} = 1$\,nA.}
		\label{Fig:high-resol}
\end{figure}
Finally, for very thick Gd films we observe winding stripe domains which are no longer aligned along a preferred direction. 
Especially in Fig.\,\ref{Fig:thickness_p2}(j) branching domains can be identified. 
Interestingly, this now includes both, dark and bright irregular stripe domains 
which clearly present weaker and rather narrow bright and dark intermediate d$I$/d$U$ signals, respectively.  
This observation suggests that the above-mentioned absence of intermediate maxima along some stripe direction in thinner films 
was indeed caused by a particular tip magnetization direction unsuitable for resolving these closure domains. 

As further evidence for this hypothesis we present the data of Fig.\,\ref{Fig:high-resol} 
which were obtained on the same 400\,AL Gd film.  
As described in Sec.\,\ref{sect:Experiment}, the W tip was initially prepared by poking the tip into the Gd film. 
This Gd/W tip was then used to measure the magnetically sensitive d$I$/d$U$ map presented in Fig.\,\ref{Fig:high-resol}(a). 
Afterwards, the tip was again dipped into the Gd film to modify its quantization axis. 
This tip was then used to obtain the data of Fig.\,\ref{Fig:high-resol}(b).  
Although the data were measured on the same Gd film and the observed domain structures are quite similar 
as far as the stripe periodicity and the general appearance are concerned, some details are strikingly different.  
In particular, we notice that the intermediate contrast can clearly be recognized 
for the entire data set of Fig.\,\ref{Fig:high-resol}(a), i.e., irrespective of the stripe orientation.
In contrast, the intermediate contrast can only be found for one stripe direction in Fig.\,\ref{Fig:high-resol}(b).
These data evidence that the apparent absence of branching domains observed in some images, 
cf.\ Fig.\,\ref{Fig:thickness_p2}(d,f,h), is not caused by their factual inexistence, 
but rather by an unfavorable orientation of the tip magnetization, rendering the observation impossible. 

\section{Summary}
\label{subsec:Summary}
\vspace{-0.3cm}
In this contribution we presented a detailed spin-polarized scanning tunneling 
microscopy (SP-STM) study of the epitaxial Gd films grown on a W(110) surface.  
To the best of our knowledge, our data provide the first high-spatial resolution images of the magnetic domains structure of this rare-earth metal. 
Great care was taken to identify optimal preparation conditions which produce the lowest density of surface defects.  
Best results were obtained for room temperature deposition followed by subsequent annealing at 900\,K.  
Whereas lower or higher annealing temperatures led to domain wall pinning or discontinuous films, respectively, 
optimally prepared Gd films with a thickness of 30 atomic layers (AL) exhibit large in-plane magnetized domains with 
domain walls which are approximately oriented along the $[1{\bar 1}0]$ direction of the underlying W substrate.
In good agreement with earlier spatially averaging studies \cite{PhysRevB.50.6457,BergerJMMM,PhysRevB.52.1078}, 
we find a spin reorientation transition (SRT) at a critical film thicknesses $\Theta_{\text{crit}} \approx (100 \pm 20)$\,AL 
where the easy axis of magnetization rotates from parallel to perpendicular to the film plane.  
SP-STM images taken in the thickness-regime of the SRT reveal a magnetic domain structure 
which appears to be spatially inhomogeneous, with patches showing a stripe domain phase 
that coexist with paramagnetic or in-plane ferromagnetic regions.  
At the coverages exceeds $\Theta_{\text{crit}}$ we find stripe domains 
which are rotated by $\approx \pm 30^{\circ}$ with respect to the W$[001]$ direction.  
The periodicity of the stripe domains is investigated thoroughly.   
Furthermore, at $\Theta \gtrsim 200$\,AL we find a magnetic domains structure which resembles a zig-zag pattern.  
Irregular stripe domains are imaged beyond 500\,AL.  
Our experimental data can consistently be explained by the interplay between different contribution to the total energy, 
i.e., the magneto-crystalline, the magneto-static, and the magneto-elastic energy density.

\section*{Acknowledgments}
\label{sect:Acknowledgments}
\vspace{-0.3cm}
This work was supported by the DFG through SFB 1170 (project A02).  
We also acknowledge financial support by the Deutsche Forschungsgemeinschaft (DFG, German Research Foundation) 
under Germany's Excellence Strategy through W{\"u}rzburg-Dresden Cluster of Excellence 
on Complexity and Topology in Quantum Matter -- ct.qmat (EXC 2147, project-id 390858490).

\end{document}




\title{Supplementary Information for: \\
``Magnetic domain structure of epitaxial Gd films grown on W(110)''}	

\author{Patrick Härtl}
	\email{patrick.haertl@physik.uni-wuerzburg.de} 
	\address{Physikalisches Institut, Experimentelle Physik II, 
	Universit\"{a}t W\"{u}rzburg, Am Hubland, 97074 W\"{u}rzburg, Germany}
\author{Markus Leisegang}
	\affiliation{Physikalisches Institut, Experimentelle Physik II, 
	Universit\"{a}t W\"{u}rzburg, Am Hubland, 97074 W\"{u}rzburg, Germany} 
\author{Matthias Bode} 
	\address{Physikalisches Institut, Experimentelle Physik II, 
	Universit\"{a}t W\"{u}rzburg, Am Hubland, 97074 W\"{u}rzburg, Germany}	
	\address{Wilhelm Conrad R{\"o}ntgen-Center for Complex Material Systems (RCCM), 
	Universit\"{a}t W\"{u}rzburg, Am Hubland, 97074 W\"{u}rzburg, Germany}    

\date{\today}
\pacs{}
\maketitle

\newpage
\section{I. Evaluation of the deposition rate}

\begin{figure}[htbp]
	\includegraphics[width=\linewidth]{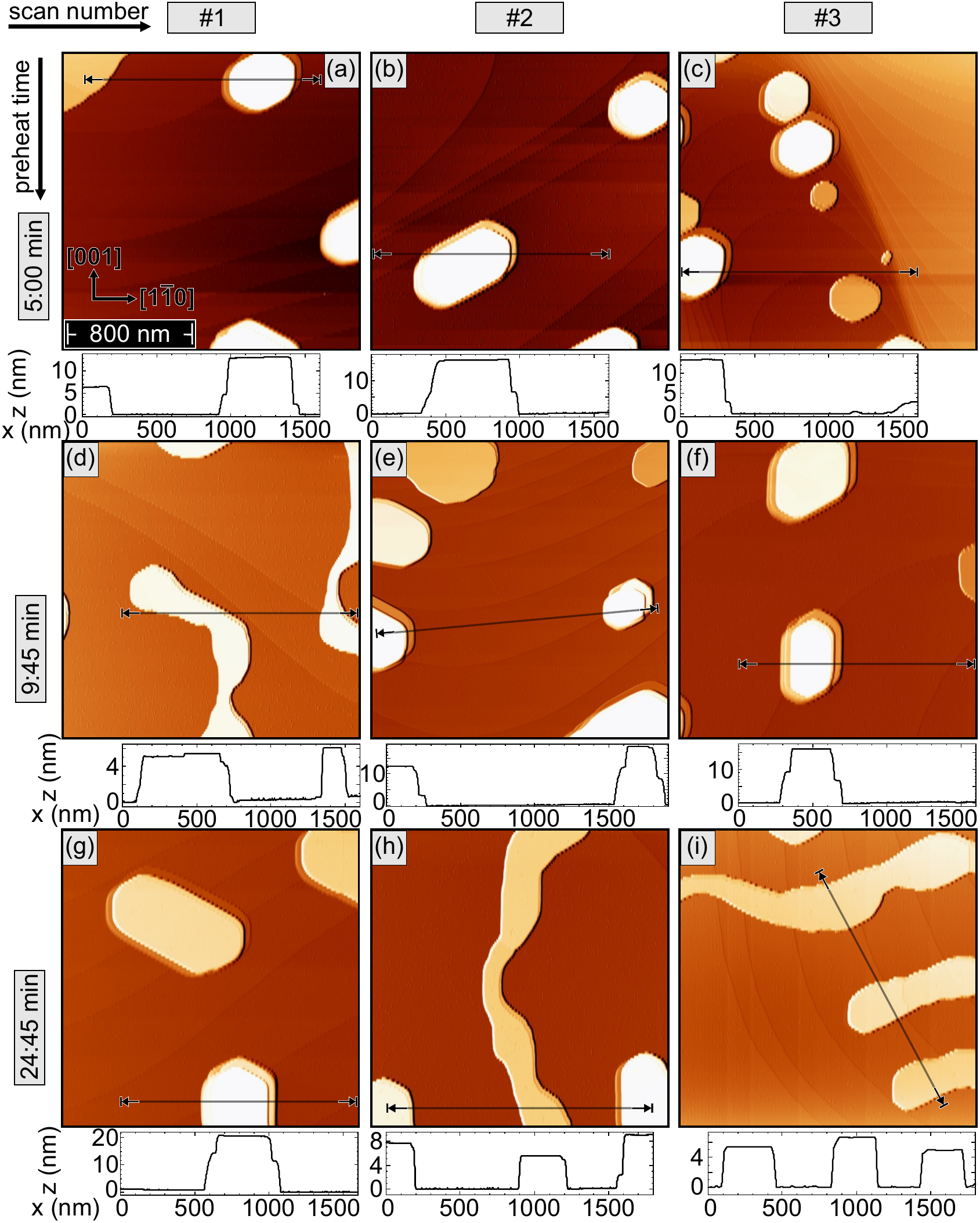}%
		\caption{Gd films grown on W(110) at different preheating times of the evaporator (see rows). 
		For each sample, three scans in different regions of the sample were taken (columns \#1, \#2, and \#3). 
		Each sample was post-annealed for five minutes at ${T>900}$\,K, resulting in Stranski-Krastanov growth. Scan parameters: ${U_{\text{bias}} = -700\,\text{mV}}$ and ${I_{\text{set}} = 500\,\text{pA}}$.}
	\label{Fig:dep-rate} 
\end{figure}

Figure~\ref{Fig:dep-rate} summarizes nine topographic scans of 5\,AL Gd films on W(110), taken to determine the deposition rate of the Gd-evaporator. 
Gd films were prepared with different preheating times, followed by a 15\,s deposition 
and a consecutive five minute post-annealing on an e-beam-stage, resulting in Stranski-Krastanov growth. 
The rows in Fig.\,\ref{Fig:dep-rate} represent experiments performed with different preheating times
i.e., 5:00\,min in Figs.\,\ref{Fig:dep-rate}(a,b,c), 9:45\,min in Figs.\,\ref{Fig:dep-rate}(d,e,f), 
and 24:45\,min in Figs.\,\ref{Fig:dep-rate}(g,h,i). 
The different preheating times were chosen to exclude potential effects of a non-linear deposition rate.

The three different preparations (nine scans) show more or less the same behavior. 
The W(110) is covered by a wetting layer of Gd followed by large islands 
with a diameter up to several hundred nanometer and heights up to 20\,nm. 
At the upper right side of high islands a multiple-tip feature can be found 
which was taken into account when determining the deposition rate.
The coverages obtained for the nine scans are summarized in Table\,\ref{tab:dep-rate}. 
The values vary around 18-20\,nm with two anomalies for scans \#1 
with preheat durations $9{:}45$\,min (12.38\,AL) and $24{:}45$\,min (27.30\,AL). The reason for that is most likely the non-homogeneous coverage of the whole sample during deposition. Overall, we determine the mean value and the standard deviation of the deposition rate to ${(19.75\pm 3.66)}$\,AL/min.

\begingroup
\begin{table}[b]
	\setlength{\tabcolsep}{20pt}
	\renewcommand*{\arraystretch}{1.0}
	\centering
	\begin{tabular}{|c|c|c|c|}
		\hline
		$\bm {scan~\#}$  & $\bm{5{:}00\,{\textbf{min.}}}$ & $\bm{9{:}45\,{\textbf{min.}}}$ & $\bm{24{:}45\,{\textbf{min.}}}$ \\
		\hline
		1	& 18.45 & 12.38 & 27.30 \\
		\hline
		2	& 21.38 & 20.19 & 18.71 \\
		\hline
		3	& 18.17 & 20.66 & 20.43 \\
		\hline
		average	& 19.36 & 16.28 & 22.15 \\
		\hline
	\end{tabular}
	\vspace*{10pt}
	\caption{Deposition rates in AL/min obtained by analyzing the data presented in Fig.\,\ref{Fig:dep-rate},
	i.e., on three different samples for three regions each.  
	Both the preheating time and the positions was modified but had no significant effect.  
	The overall deposition rate is determined to ${(19.75\pm 3.66)}$\,AL/min.}
	\label{tab:dep-rate}
	\end{table}
\endgroup

\newpage
\section{II. Data Processing}
\begin{figure}[htbp]
	\includegraphics[width=0.93\linewidth]{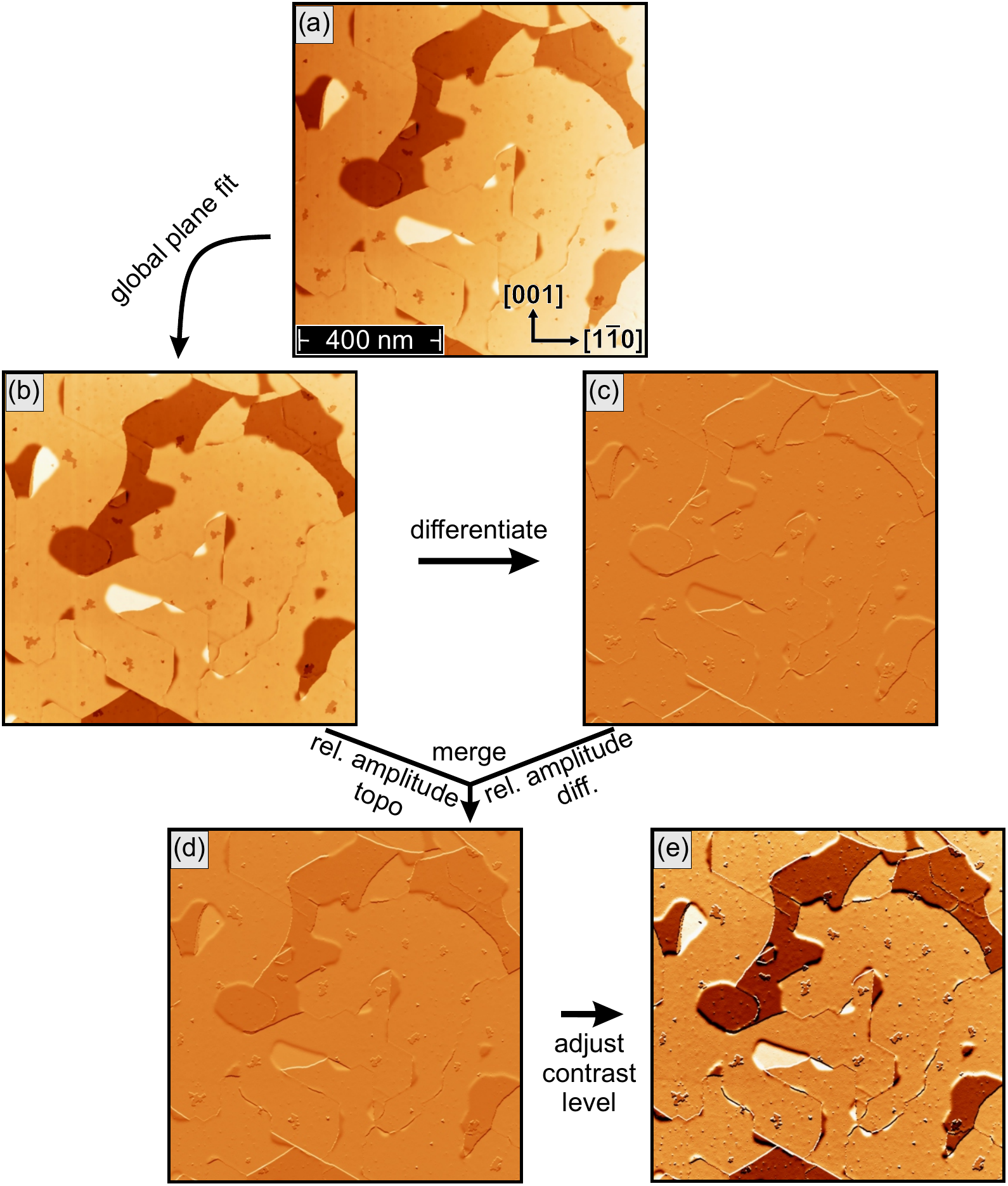}%
		\caption{Schematic process of the topographic data processing used in this work. 
			The process is exemplarily shown for the 30\,AL Gd film shown in the main text in Fig.\,3(b), 
			see main text for preparation details. 
			Scan parameters: ${U_{\text{bias}} = -700\,\text{mV}}$ and ${I_{\text{set}} = 1\,\text{nA}}$.} 
	\label{Fig:model} 
\end{figure}

The topographic as well as the magnetic d$I$/d$U$ data were post-processed with WSxM, 
a software especially made for scanning probe microscopy data \cite{horcas2007wsxm}. 
As for the magnetic d$I$/d$U$ data only the contrast was adjusted, 
the topographic data needed a bit more processing to point out defect structures in more detail. 
The procedure was as followed:
\begin{enumerate}
\item Figure~\ref{Fig:model}(a) shows the raw original data, as measured.
\item With a global plane fit, the overall surface is flattened, see Fig.\,\ref{Fig:model}(b).
\item The plane fitted data is then differentiated in Fig.\,\ref{Fig:model}(c).
\item The two former pictures are then merged, to enhance structural defects. 
		Both relative amplitudes are chosen such, that the height information still remains visible. 
		Typically, the topography is added with a factor 50:1 to the differentiated image, see Fig.\,\ref{Fig:model}(d).
\item Finally, the contrast is adjusted to emphasize out the details. 
		The final picture is shown in Fig.\,\ref{Fig:model}(e).
\end{enumerate}

\newpage
\section{III. Preparation of an unpolarized W tip on Ag(111)}
\begin{figure}[htp]
	\includegraphics[width=0.867\linewidth]{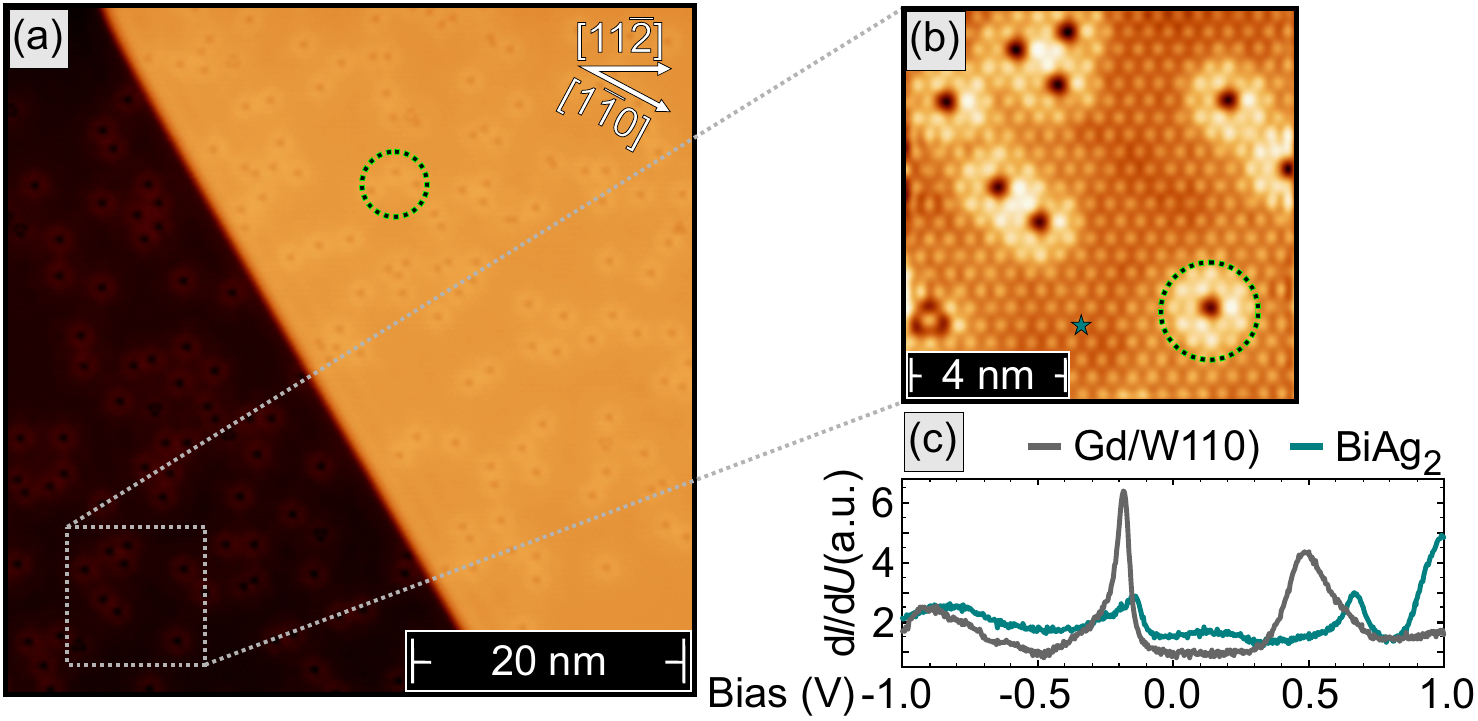}%
		\caption{(a) Overview scan of a Bi/Ag(111) alloy surface, with a zoom-in in (b) for the preparation of an unpolarized tip. 
		(c) d$I$/d$U$ point spectrum of the Rashba-split surface state of the BiAg$_{2}$ alloy 
		at $eU = -124$\,meV and $eU = +704$\,meV (green line), taken at the position of the green star, marked in (b). 
		Additionally, the Gd spectrum taken afterwards on a 200\,AL thick Gd film grown on W(110) 
		as a reference to exclude a tip change, is shown with the grey line (details see main text). 
		Scan parameters: ${U_{\text{bias}} = -50\,\text{mV}}$ and ${I_{\text{set}} = 1\,\text{nA}}$. }
	\label{Fig:BiAg2} 
\end{figure}

Figure~\ref{Fig:BiAg2} summarizes the preparation of an unpolarized W tip on a Bi/Ag(111) surface alloy. 
Therefore the Ag(111) crystal was cleaned by two consecutive cycles of sputtering (30\,min, 500\,eV, $2.0\times 10^{-6}$\,mbar) 
and annealing (20\,min, ${T_{\textrm{anneal}}\approx 720\,\textrm{K}}$), 
with a consecutive deposition of $\approx 1/3$\,AL Bismuth (Bi) 
onto the warm sample (${T_{\textrm{deposit}}\approx 570}$\,K) leading to a BiAg$_2$ alloy. 
This preparation is shown in Fig.\,\ref{Fig:BiAg2}(a). 
Since the amount of Bi in this preparation was just below $1/3$\,AL, 
resulting in vacancies in the BiAg$_{2}$ alloy, as exemplarily marked by green dashed circles in Figs.\,\ref{Fig:BiAg2}(a,b). 
The zoom-in in Fig.\,\ref{Fig:BiAg2}(b) reveals defect-free patches of several nanometers, 
allowing to measure its surface state at the position marked by the green star. 

The BiAg$_2$ surface alloy is formed by replacing every third Ag atom by a Bi atom, 
resulting in a ${(\sqrt{3}\times\sqrt{3}})\textrm{Bi}/\textrm{Ag}(111)R30^{\circ}$ surface alloy 
with Rashba-split surface states, see d$I$/d$U$ spectrum in Fig.\,\ref{Fig:BiAg2}(c) - green line. 
The occupied part is measured at a binding energy of $eU = -124$\,meV, while the unoccupied part is $eU = +704$\,meV. 
Except of the sharp increase above $U_{\textrm{bias}} = +800$\,meV 
and a slight elevation at around $U_{\textrm{bias}}= - 900$\,meV, 
no other significant features in this d$I$/d$U$ signal can be identified.

After characterization of the unpolarized W tip on the BiAg$_2$ alloy, including scans over step edges, thereby excluding double-tip features and a spectrum without any distinct tip-states, a 200\,AL thick Gd film was prepared (details see main text, section \textbf{B. Electronic Properties} and Fig.\,4). 
The d$I$/d$U$ maps in Fig.\,4 revealed an almost flat signal, indicating no magnetic contrast, 
as is also supported by the d$I$/d$U$ spectrum on the Gd/W(110) surface, 
showing no ``magnetic'' features in the spectrum [see grey line in Fig.\,\ref{Fig:BiAg2}(c)], 
only the Gd(0001) split surface state at $eU = -190$\,meV and $eU = +480$\,meV (see main text for more details) 
and the slight elevation at are visible $\approx U_{\textrm{bias}} = - 900$\,meV can be observed. 
This leads to the conclusion, that the tip still shows unpolarized behavior.

\clearpage
\section{IV. Verification of same tip}
\begin{figure}[hb]
	\includegraphics[width=0.75\linewidth]{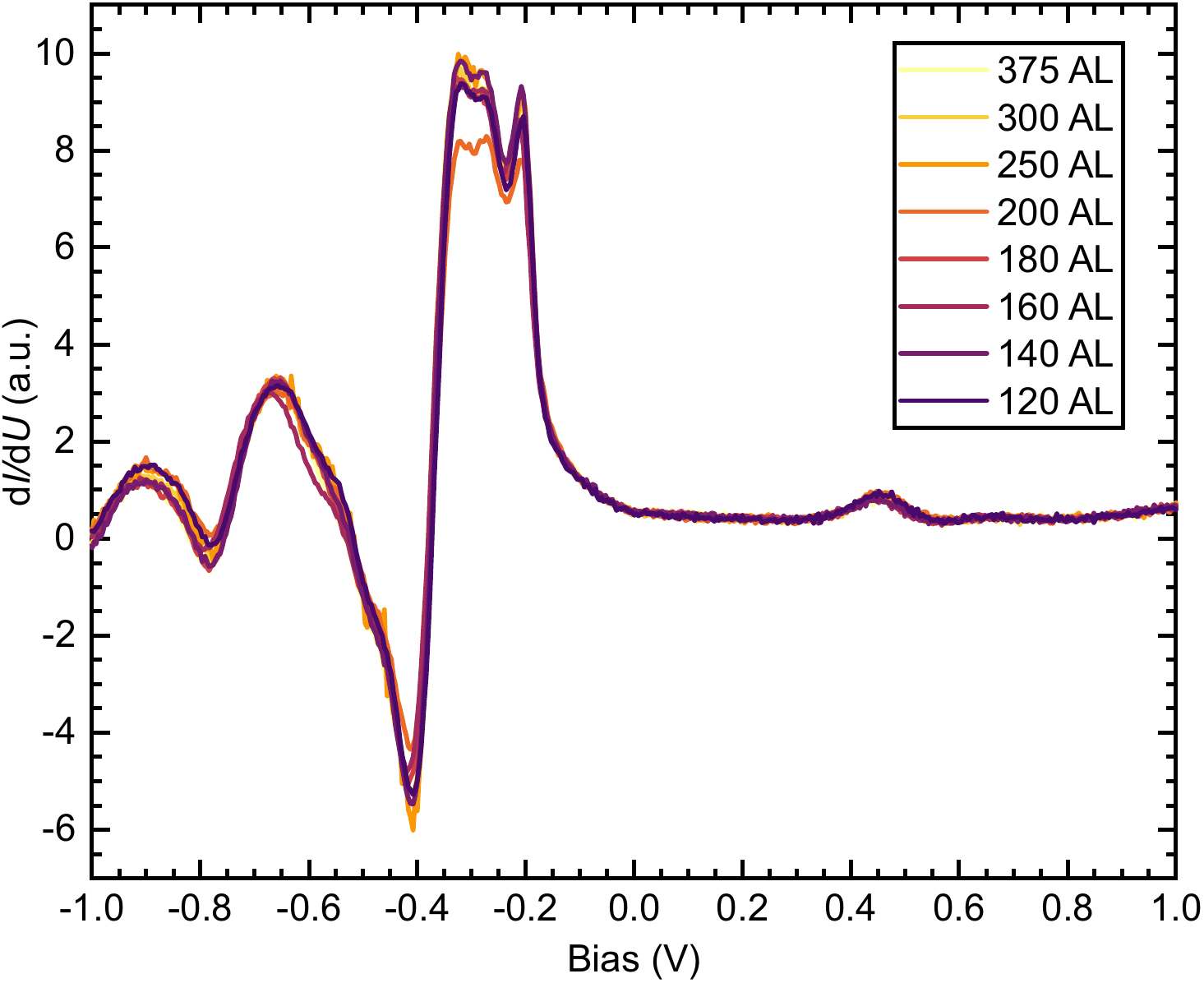}%
		\caption{Summary of all d$I$/d$U$ point spectra taken on a film coverage dependent series with the same magnetically sensitive tip from Figs.\,6 and 7 in the main text. The spectra were taken on comparable areas of each sample in the region of magnetically dark contrast for film thicknesses of 120, 140, 160, 180, 200, 250, 300 and 375\,AL. The spectra show alongside the surface states at $eU = -190$\,meV and $eU = +480$\,meV a strong tip state around $eU \approx -300$\,meV down to $eU \approx -600$\,meV, causing the stark magnetic contrast in the d$I$/d$U$ maps in Figs.\,6 and 7 in the main text.}
	\label{Fig:sametip} 
\end{figure}

Fig.\ref{Fig:sametip} summarizes point-spectra taken on some samples for the coverage-dependent study of Figs.\,6 and 7. The point spectra were taken on each sample in regions of dark magnetically contrast. The tip remained the same for film thicknesses of 120, 140, 160, 180, 200, 250, 300 and 375\,AL, as can be seen in the figure. Besides the surface states of Gd at $eU = -190$\,meV and $eU = +480$\,meV, a strong tip state at $eU \approx -300$\,meV down to $eU \approx -600$\,meV is present, giving rise for the strong contrast in the coverage dependent study in Figs.\,6 and 7 in the main text.

\newpage
\section{V. Stripes in the different direction}
\begin{figure}[hb]
	\includegraphics[width=\linewidth]{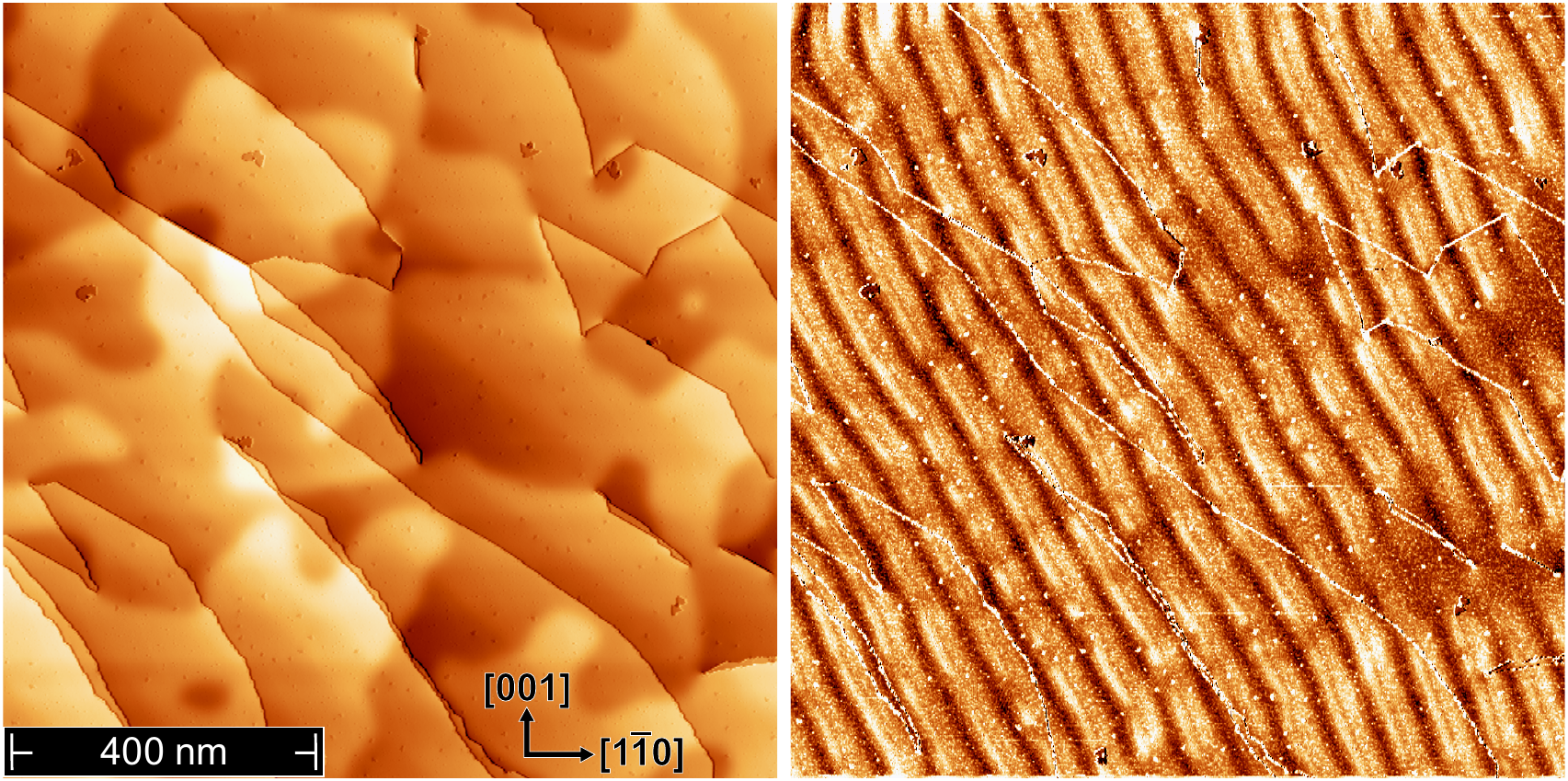}%
		\caption{Overview scan of a 140\,AL thick Gd-Film epitaxially grown on the W(110) substrate. The topography in (a) shows a typical topography for an optimally prepared film (see main text for preparation details). The magnetically sensitive d$I$/d$U$ signal now show stripes rotated $-30^{\circ}$ against the W[001] direction, compared to the stripes in Figs.\,6(h, j, l), where the stripes are rotated in the $+30^{\circ}$ direction. Scan parameters: ${U_{\text{bias}} = -700\,\text{mV}}$ and ${I_{\text{set}} = 1\,\text{nA}}$.}
	\label{Fig:stripe-direction} 
\end{figure}

Fig.\,\ref{Fig:stripe-direction} summarizes the topography (a) and the simultaneously recorded d$I$/d$U$ signal (b) for an optimally prepared 140\,AL thick Gd(0001)/W(110) film. Preparation details see main text. The topography in Fig.\,\ref{Fig:stripe-direction}(a) displays a typical topography for such a Gd-film with double-screw dislocations, smoothed steps, point-like indentations and some hydrogen contaminations, due to the not perfectly clean gadolinium. The magnetically sensitive d$I$/d$U$ signal in Fig.\,\ref{Fig:stripe-direction}(b) reveals now stripe domains rotated $-30^{\circ}$ against the W[001] direction, as it is the opposite direction of the coverage dependent study in Figs.\,6 and 7 in the main text.

\newpage
\section{VI. Bias dependent contrast}

\begin{figure}[hb]
	\includegraphics[width=\linewidth]{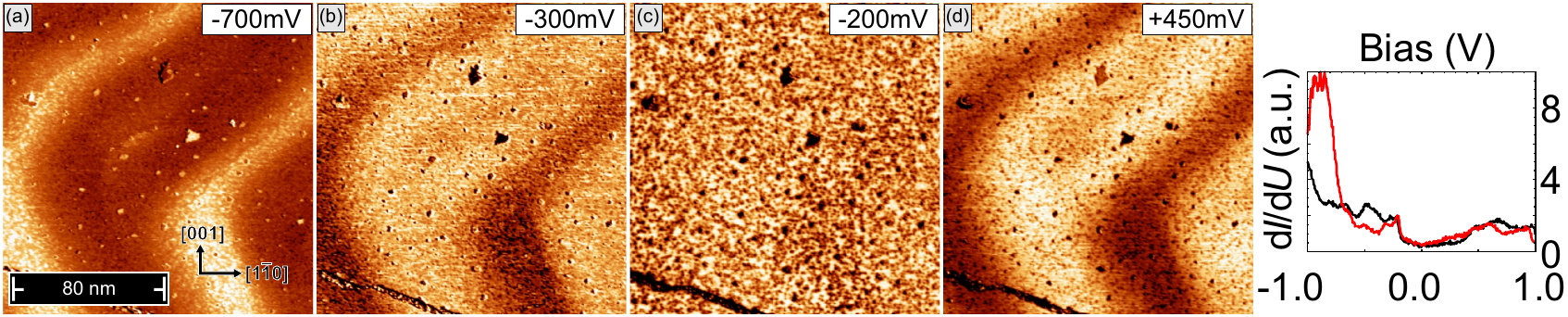}%
		\caption{Bias dependent contrast}
	\label{Fig:bias-dependent} 
\end{figure}


\newpage